\documentclass[useAMS]{mn2e}
\usepackage{psfig,epsfig}
\usepackage{graphicx}
\usepackage{amssymb}
\usepackage{epstopdf}

\begin{document}
\def\simlt{\mathrel{\rlap{\lower 3pt\hbox{$\sim$}}
        \raise 2.0pt\hbox{$<$}}}
\def\simgt{\mathrel{\rlap{\lower 3pt\hbox{$\sim$}}
        \raise 2.0pt\hbox{$>$}}}

\title[The PEP survey: star-forming activity in radio-AGN]{The PEP Survey: evidence for intense star-forming activity in the majority of radio-selected AGN at $z\simgt 1$.}

\author[Manuela Magliocchetti et al.]
{\parbox[t]\textwidth{M. Magliocchetti$^{1}$,  D. Lutz$^{2}$, P. Santini$^{3}$, M. Salvato$^{2}$, P. Popesso$^{4}$, S. Berta$^{2}$, F. Pozzi$^{5}$\\
} \\
{\tt $^1$ INAF-IAPS, Via Fosso del Cavaliere 100, 00133 Roma,
  Italy}\\
  {\tt $^2$ Max Planck Institut f\"ur extraterrestrische Physik (MPE),
  Postfach 1312,  D85741, Garching, Germany}\\
  {\tt $^3$ INAF-Osservatorio Astronomico di Roma, Via di Frascati 33, 00040 Monte Porzio Catone, Italy}\\
  {\tt $^4$ Excellence Cluster, Boltzmannstr. 2, D85748, Garching, Germany}\\
  {\tt $^5$ Dipartimento di Astronomia, Universita' di Bologna, Via Ranzani 1, 40127, Bologna, Italy}\\
    }
  \maketitle
\begin{abstract}
In order to investigate  the FIR properties of  radio-active AGN, we have considered three different fields where both radio and FIR observations are the deepest to-date: GOODS-South, GOODS-North and the Lockman Hole. Out of a total of 92 radio-selected AGN, $\sim 64\%$ are found to have a counterpart in {\it Herschel} maps. The percentage is maximum in the GOODS-North ($72$\%) and minimum ($\sim 50$\%) in the Lockman Hole, where FIR observations are shallower. Our study shows that in all cases FIR emission is  associated to star-forming activity within the host galaxy. Such an activity can even be extremely intense, with 
star-forming rates as high as $\sim 10^3-10^4$ M$_\odot$yr$^{-1}$. 
AGN activity does not inhibit star formation in the host galaxy, just as on-site star-formation does not seem to affect AGN properties,  at least those detected at radio wavelengths and for $z\simgt 1$. 
Given the very high rate of FIR detections, 
we stress that this refers to the majority of the sample: most radio-active AGN are associated with  intense episodes of star-formation. However, the two processes proceed independently within the same galaxy, at all redshifts but in the local universe, where powerful enough radio activity reaches the necessary strength to switch off the on-site star formation.  
Our data also show that for z $\simgt 1$ the hosts of radio-selected star-forming galaxies and AGN are indistinguishable from each other both in terms of mass and IR luminosity distributions. The two populations only differentiate in the very local universe, whereby the few AGN which are still FIR-active  are found in galaxies with much higher masses and luminosities.
\end{abstract}

\section{introduction}

Radio sources constitute an unequally powerful tool to investigate the origin and evolution of our Universe. In fact, radio emission is not obscured or attenuated by dust or gas, so deep radio surveys offer a unique opportunity to detect and study sources up to the highest redshifts.   It is following this rationale that very ambitious programs like the planned Square Kilometer Array (SKA, Carilli et al., 2004) facility, a gigantic radio telescope with a total collective area of 1 square kilometer, or its precursors ASKAP (Australian SKA pathfinder, Johnston S. et al. 2007) and MeerKAT (Jonas J.L. 2009) will soon see their first light. When applied to radio sources, the expected advantages in sensitivity, field-of-view, frequency range and spectral resolution will yield substantial progress in many research fields, from cosmology to astrophysics.

Extra-galactic radio sources are however a mixed bag of astrophysical objects:  amongst the most powerful ones we find radio-loud QSOs and FRII (Fanaroff \& Riley 1974) galaxies while, moving to weaker sources, the dominant populations become FRI,  low-excitation galaxies and  star-forming galaxies, whose contribution to the total radio counts become predominant at the sub-mJy level (see e.g. Magliocchetti et al. 2000; Prandoni et al. 2001).
Discerning amongst these objects and their relative weight is an impossible task to carry out only on the basis of radio-continuum data. This is why increasingly more effort was recently put on photometric and spectroscopic follow-up studies of radio sources detected in deep-field surveys (amongst the many: Schinnerer et al. 2004; 2007; 2010 for the COSMOS field;  Morrison et al. 2010 for GOODS-N; Mc Alpine et al. 2013 for the VIDEO-XMM3 field;  Bondi et al. 2003 for the VVDS field: Simpson et al. 2012 for the Subaru/XMM-Newton Deep Field: Mao et al. 2011 for the Extended Chandra Deep Field). 
Investigations of the optical properties of radio sources can then disentangle the different populations and also allow the determination of a number of quantities which influence the source behavior like e.g. the black-hole mass in optical spectra dominated by broad emission lines (e.g. Metcalf \& Magliocchetti 2006).   

Following the launch of the {\it Spitzer} satellite, in recent years radio sources have also been investigated at Mid-Infrared (MIR) wavelengths (e.g. Appleton et al. 2004; Boyle et al. 2007; Magliocchetti, Andreani \& Zwaan 2008; Garn et al. 2009;  Leipski et al. 2009; De Breuck et al. 2010; Norris et al. 2011 amongst the many). These studies proved to be extremely useful for two main reasons: they allowed  investigations of the properties and evolution of the sub-population of star-forming galaxies up to $z\sim 2$, while on the other hand they  provided interesting hints on the  central engine responsible for radio (and MIR) AGN emission.

It is only in these very last few years that the radio community has started to realize the importance of multi-wavelength studies which also include Far-Infrared (FIR) information. This growing interest is primarily due to the advent and launch of the {\it Herschel} satellite, which for the first time has observed galaxies at FIR wavelengths up to very large cosmological ($z\sim 4$) distances (Lutz 2014). Earlier FIR missions such as IRAS, ISO and {\it Spitzer}@70$\mu$m only probed the relatively local ($z\simlt 1$) universe, and since AGN-powered radio galaxies in the local universe are almost always hosted by massive red galaxies with little or no ongoing star-formation activity (e.g. Magliocchetti et al. 2002; 2004), and since FIR emission in galaxies is almost entirely due to processes connected with ongoing star formation, it follows that any pre-{\it Herschel}  interest in a joint radio-FIR analysis was limited to re-assessing the precision of the tight radio-FIR correlation found by early IRAS-based studies of local star-forming galaxies (Condon et al. 1982).

However, when moving to higher redshifts, galaxies undergo dramatic changes.  For instance, it is now well established that both the cosmic star-forming activity and AGN output steadily increase with look-back time at least up to redshifts $z\sim2$ (e.g. Gruppioni et al. 2013: Merloni, Rudnick \& Di Matteo 2004), and many galaxies are observed to host both an active AGN and ongoing star-formation (e.g. Alexander et al. 2008). It is then legitimate to wonder whether radio galaxies also follow such a trend and if in the early universe they are also associated with ongoing stellar production. 
First attempts in this direction were presented by the work of Seymour et al. (2011) which analyses the FIR emission as observed by the SPIRE instrument (Griffin et al. 2010) on board of the {\it Herschel} satellite (Pilbratt et al. 2010) for a sample of very powerful, 
L$_{1.4 \rm GHz}\ge 10^{25}$ W Hz$^{-1}$ radio sources selected in the extra-galactic {\it Spitzer} First Look Survey field, and by the work of Del Moro et al. (2013) who investigate hidden AGN activity in a sample of GOODS-North sources with deep radio, infrared and X-ray data.

Magliocchetti et al. (2014)  pushed the Seymour et al. (2011) results further and investigated, for the first time, the FIR properties of radio-selected AGN of {\it all} radio luminosities and at {\it all} redshifts $z\simlt 3.5$. 
This was done by using the very deep radio catalogue obtained at 1.4 GHz in the COSMOS field by Schinnerer et al. 2004; 2007; 2010 and Bondi et al. (2008); AGN were selected solely on the basis of their radio-luminosity and FIR information came from the PEP survey (Lutz et al. 2011). Results from this analysis indicate that  the probability for a radio-selected AGN to be detected at FIR wavelengths is both a function of radio power and redshift. 
Powerful sources were found more likely to be FIR emitters at earlier epochs because of two distinct effects: 1) at all radio luminosities, FIR activity monotonically increases with look-back time and 2) in the earlier universe, radio activity of AGN origin is increasingly less effective at inhibiting FIR emission.  Magliocchetti et al. (2014) also found that the observed FIR (and MIR) emission was due to processes which were indistinguishable from those which power star-forming galaxies. 

The above results, although interesting and obtained on a relatively large area, however need the complementarity of  deep enough FIR observations which can match the depth of the radio data. As a matter of fact, only $\sim 36$\% of the sources identified in the COSMOS field as radio-emitting AGN possessed a counterpart in the PEP catalogues either at 100 $\mu$m or at 160 $\mu$m. 
The present work then overcomes the above limitations and analyses the FIR properties of radio-selected AGN  on three fields where FIR observations are the deepest-to-date: GOODS-North, GOODS-South and the Lockman Hole. 
FIR data come from both the PEP survey (Lutz et al. 2011) and the {\it Herschel}-GOODS Survey (Elbaz et al. 2011) and, in the best-observed areas such as the two GOODS fields, ensure limiting FIR fluxes which are deeper by as much as a factor $\sim 6$ at 100 $\mu$m and $\sim 5$ at 160 $\mu$m than the PEP observations on COSMOS.
Throughout the work we will assume a $\Lambda$CDM cosmology with $H_0=70 \: \rm km\:s^{-1}\: Mpc^{-1}$ ($h=0.7$), $\Omega_0=0.3$,  $\Omega_\Lambda=0.7$ and $\sigma^m_8=0.8$. In all cases, masses for the sources under examination are calculated by using the Bruzual \& Charlot (2003) templates and adopting  a Salpeter (1955) Initial Mass Function  (IMF).

\section{The Fields}

\subsection{Lockman Hole}

\begin{figure}
\begin{center}
\includegraphics[scale=0.42]{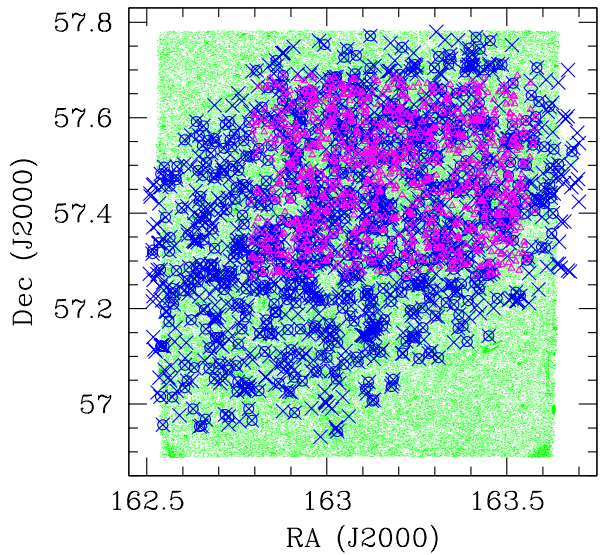}
\caption{Lockman Hole. (Blue) crosses represent radio sources selected at 1.4 GHz by Ibar et al. (2009), (green) small dots show the area covered by optical and near-IR surveys, while (magenta) open triangles illustrate the sources imaged by the PEP survey. 
Crosses surrounded by open circles indicate radio sources with a redshift (either spectroscopic or photometric) determination from the catalogue of Fotopoulou et al. (2012).
\label{fig:LH}}
\end{center}
\end{figure}

The radio sources used in this work have been taken from the paper of Ibar et al. (2009). 
For consistency with the other surveys considered in this paper, we focussed on the catalogue of objects observed with the Very Large Array (VLA) at 1.4 GHz. 
The sensitivity of these observations was $\sigma\sim 6\mu$Jy beam$^{-1}$, with a resolution (Full Width Half Maximum - FWHM) of $\sim 5$ arcsec.
 Ibar et al. (2009) provide a catalogue of 1303 unique sources (1346 including sub-components of composite radio objects), selected to have a peak-to-local noise (PNR) greater than 5. 
 As these objects were also observed at 610 MHz with the Giant Metrewave Radio Telescope (GMRT), the overwhelming majority of them is also endowed with an estimate of the radio spectral index 
 $\alpha_{1400-610}$ (hereafter just $\alpha$ for clarity, where $\alpha$ is defined as F$_\nu\propto \nu^{-\alpha}$, with F radio flux). Ibar et al. (2009) also provide 
 a classification based on the radio source appearance, and distinguish them into single (S), double (D), triple (T) and Extended (M), plus some more complicated morphological cases such as single source with one close neighbour (SD), 
 single source with multiple close neighbours (ST) and so on. Figure \ref{fig:LH} shows the radio sources as taken from the Ibar et al. (2009) 1.4 GHz selection as (blue) crosses.

Optical, near-IR and mid-IR information for these sources, as well as their redshift determination, can be obtained from the work of Fotopoulou et al. (2012) which provides very precise broad band photometry and photometric redshifts for about 188,000 sources located in 
$\sim 0.5$ deg$^2$ of the Lockman Hole area (green small dots in Figure \ref{fig:LH}). Special care was taken for objects identified as AGN. Note that the Fotopoulou et al. (2012) catalogue also includes 388 X-ray detected sources (the overwhelming majority of which are active galactic nuclei) identified in the deep XMM-Newton observations by Brunner et al. (2008), available for an area of 0.2 deg$^2$. \\
Optical identifications for the Ibar et al. (2009) radio sources were then obtained from the Fotopoulou et al. (2012) catalogue for a maximum tolerance radius of 0.9$^{\prime\prime}$. This value was chosen so to maximize the number of real associations, while reducing the percentage of spurious matches to roughly 6$\%$.  The total number of radio sources on the area covered by the broad-band observations of Fotopoulou et al. (2012) is 1309. The number of sources with a photometric id obtained in the way described above is 723, corresponding to 55,7\% of the original parent sample.

\begin{figure}
\begin{center}
\includegraphics[scale=0.42]{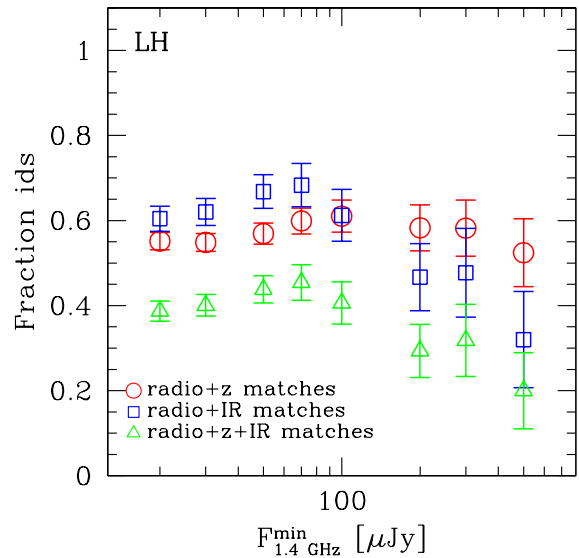}
\caption{Fraction of radio sources in the Lockman Hole endowed with a redshift determination (red circles), with a FIR counterpart from the PEP catalogue (blue squares) and with both a FIR id and a redshift determination (green triangles) as a function of integrated 1.4 GHz flux. 
Error-bars correspond to Poissonian estimates.
\label{fig:fraction_LH}}
\end{center}
\end{figure}

The fraction of radio sources in the Lockman Hole endowed with either a spectroscopic or photometric redshift determination as a function of integrated 1.4 GHz radio flux is shown in Figure \ref{fig:fraction_LH} by the red circles, where the associated error-bars correspond to Poissonian estimates. 
From the figure, one can notice that such a fraction is constant within the errors, and equal to approximately 55\% of the parent sample throughout the whole flux range probed by the Ibar et al. (2009) survey. This indicates that no radio-flux bias affects the optical identifications of the radio sources under consideration.

A smaller area, corresponding to roughly 24$^\prime\times 24^\prime$, has also been observed by the PACS instrument (Poglitsch et al. 2010)  onboard of the {\it Herschel} satellite (Pilbratt et al. 2010) as a part of the PACS Evolutionary Probe (PEP,  D. Lutz et al. 2011) Survey, aimed at  studying the properties and cosmological evolution of the infrared population up to redshifts $z\sim 3-4$. 1328 sources were observed in this field at either 100 $\mu$m or 160 $\mu$m by the PEP survey and the blind  (i.e. source extraction  performed without relying on priors) catalogues include objects down to a 3$\sigma$ confidence level respectively of $\sim3.6$ mJy and $\sim7.5$ mJy. These are shown in Figure \ref{fig:LH} by the (magenta) open triangles.  698 radio sources from the Ibar et al. (2009) catalogue reside within the area covered by the PEP observations,  corresponding to 53,8\% of the original sample.\\ 

 In order to provide the radio sources  with a FIR counterpart, we followed the approach presented in Magliocchetti et al. (2014) and cross-matched the radio and FIR catalogues with a maximum coherence radius of 4 arcsecs. This choice insures a negligible contribution from spurious associations ($\sim4$\%). The number of radio sources with a FIR counterpart at either 100 $\mu$m or 160 $\mu$m is 422, corresponding to 60\% of the parent radio catalogue. The blue squares in Figure \ref{fig:fraction_LH} represent the trend for FIR identifications as a function of 1.4 GHz radio flux. This percentage is roughly constant up to about 100$\mu$Jy and then slowly starts declining beyond that value. The green triangles in Figure \ref{fig:fraction_LH}  represent the fraction of radio sources which are endowed with both a redshift determination and a FIR counterpart. Such a fraction is roughly equal to 40\% at the lowest fluxes and then also starts declining beyond radio fluxes F$_{1.4 \rm GHz}\simgt 100 \mu$Jy. 

\subsection{GOODS South}

\begin{figure}
\begin{center}
\includegraphics[scale=0.42]{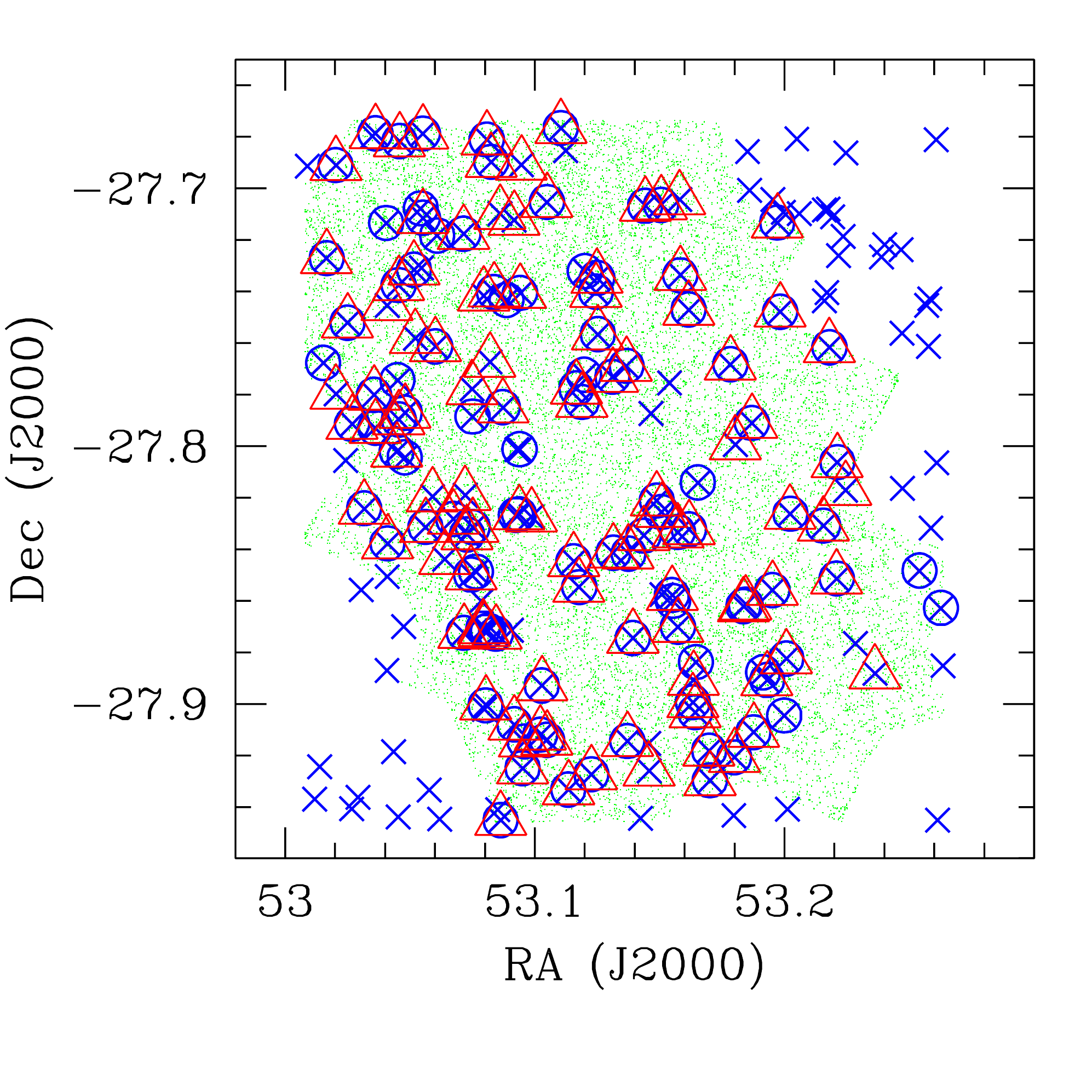}
\caption{GOODS South. (Blue) crosses represent radio sources selected at 1.4 GHz by Miller  et al. (2013), (green) small dots show the area covered by optical and near-IR surveys, while (red) open triangles illustrate the sources imaged by the PEP and the GOODS-Herschel surveys. 
Crosses surrounded by open circles indicate radio sources with a redshift (either spectroscopic or photometric) determination from the catalogue of Santini et al. (2009).
\label{fig:GS}}
\end{center}
\end{figure}

Deep, 1.4 GHz, radio observations of the whole Extended Chandra Deep Field South (ECDFS) are presented in Miller et al. (2013). This survey covers about a third of a square degree and images 883 radio sources -- out of which 17 are likely multiple-component objects -- at the 5$\sigma$ level down to a peak rms sensitivity of 6 $\mu$Jy.

A smaller area of the ECDFS, corresponding to what is called the GOODS Southern Field (hereafter GOODS-S), is one of the best studied regions in cosmology. Thanks to the very deep and exquisite observations taken at all wavelengths, Santini et al. (2009) were able to provide a high quality photometric catalogue which covers all available wavelengths from 0.3 to 24 micron.  The GOODS-MUSIC catalogue (Santini et al. 2009) also includes high precision photometric redshifts for about 15,000 objects and, whenever possible, spectroscopic redshifts for about 3000 of them. 

We then cross-correlated the  Miller et al. (2013) radio catalogue with the GOODS-MUSIC one in order to provide the 1.4 GHz-selected radio sources with a redshift determination. 142 radio objects out of the the 883 included in the Miller et al. (2013) catalogue fall within the region covered by the GOODS-MUSIC observations. 
The distribution of the Miller et al. (2013) 1.4 GHz-selected sources on the GOODS-S field is shown in Figure \ref{fig:GS} by the (blue) crosses. The green-shaded area instead shows the distribution of sources from the GOODS-MUSIC catalogue.

\begin{figure}
\begin{center}
\includegraphics[scale=0.42]{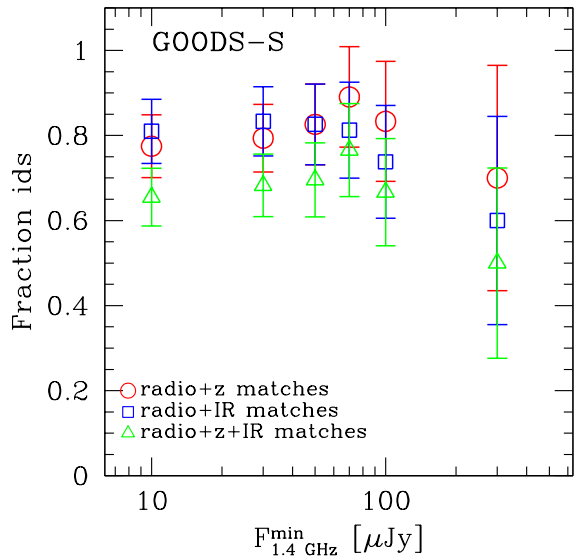}
\caption{Fraction of radio sources in the GOODS-S endowed with a redshift determination (red circles), with a FIR counterpart from the PEP catalogue (blue squares) and with both a FIR id and a redshift determination (green triangles) as a function of integrated 1.4 GHz flux. 
Error-bars correspond to Poissonian estimates.
\label{fig:fraction_GS}}
\end{center}
\end{figure}

Once again with the aim of maximizing the number of sources with a redshift determination, while at the same time minimizing the number of spurious matches, we chose the value of 0.8$^{\prime\prime}$ as the maximum radius to be considered for radio-to-optical associations. This procedure leads to 111 photometrically identified radio sources, corresponding to $\sim78$\% of the parent catalogue. The percentage of expected spurious associations is limited to a mere 3.5\%. \\
The fraction of radio sources in the GOODS-S endowed with a redshift determination as a function of integrated 1.4 GHz radio flux is shown in Figure \ref{fig:fraction_GS} by the red circles, where the associated error-bars correspond to Poissonian estimates. 
From the figure, one can notice that such a fraction is constant within the errors, and equal approximately to 80\% of the parent sample throughout the whole flux range probed by the Miller et al. (2013) survey. As it was in the case of the Lockman-Hole, the above trend proves that no radio-flux bias affects the optical identifications of the radio sources in exam.

Far infrared information for the GOODS-S radio sources can be obtained through the PEP (Lutz et al. 2011) and  PEP/GOODS-H catalogues (Magnelli et al. 2013). This last one combines observations of the GOODS fields from the PEP and GOODS-{\it Herschel} 
(Elbaz et al. 2011) key programs. This catalogue reaches 3$\sigma$ depth of 0.9, 0.6 and 1.3 mJy respectively at 70, 100 and 160 $\mu$m and includes 1635 sources. As the original PEP catalogue for GOODS-S covers a wider area than that of PEP/GOODS-H, we have identified the FIR counterparts in two steps, first by using the PEP catalogue, and then the PEP/GOODS-H one. As already specified in the previous section, the maximum tolerance radius chosen to associate a radio source to a FIR counterpart was 4$^{\prime\prime}$. 
This procedure leads to 115, 1.4 GHz-selected, radio sources with a FIR counterpart (corresponding to $\sim81$\% of the parent catalogue), out of which 110 come from the PEP catalogues either at 100 or at 160 $\mu$m and 5 from the combined PEP/GOODS-H catalogue.  The expected fraction of random associations is less than 2\%.\\

The trend for the fraction of FIR identifications as a function of integrated radio flux is presented in Figure \ref{fig:fraction_GS} by the blue squares, while that for sources with both a FIR counterpart and a redshift determination by the green triangles. Errors correspond to Poissonian estimates. Note that in both cases the fractions tend to stay constant, equal to the values of $\sim80$\% and $\sim70$\% up to radio fluxes F$_{1.4\rm GHz}\sim 100 \mu$Jy, then they start to monotonically decline.

\subsection{Goods North}

\begin{figure}
\begin{center}
\includegraphics[scale=0.42]{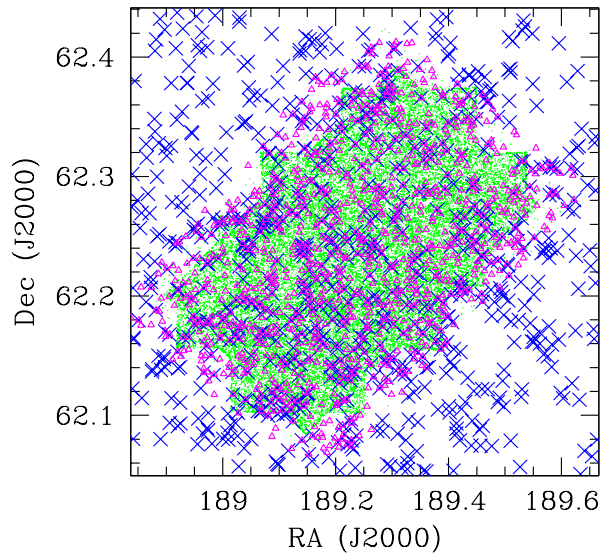}
\caption{GOODS North. (Blue) crosses represent radio sources selected at 1.4 GHz by Morrison et al. (2010), (green) small dots show the area covered by optical and near-IR surveys, while (magenta) open triangles illustrate the sources imaged by the PEP and the GOODS-Herschel surveys. 
\label{fig:GN}}
\end{center}
\end{figure}

Deep, 1.4 GHz, radio observations of the GOODS-North field (hereafter GOODS-N) are presented in Morrison et al. (2010). These authors provide a catalogue of 1230 discrete radio sources, imaged over an area of 40$^\prime \times 40^\prime$. The catalogue comprises all sources above a 5$\sigma$ detection threshold of $\sim20$ $\mu$Jy at the field centre. \\
The GOODS-N, likewise the GOODS-S, is provided with extremely deep observations at all wavelengths. This enables to derive photometric redshifts for the majority of the sources residing there. Furthermore, many spectroscopic redshifts are also available, and the PEP collaboration (Berta et al. 2011 \footnote{The multiwavelength GOODS-N catalogue is available on the PEP web page at the URL http://www.mpe.mpg.de/ir/Research/PEP/public$\_$data$\_$releases.php}) provides a list of sources observed in the optical and NIR bands, the majority of which are listed with a redshift, stellar mass and age determination. 
We then cross-correlated the GOODS-N radio-catalogue with that of optically-selected sources presented above. The number of radio objects which fall within the area observed at the other wavebands is 401. Out of these, 295 have a counterpart in the photo-z catalogue, out to a maximum radius of 0.7 arcsecs. 267 of them, corresponding to $\sim 67\%$ of the parent radio catalogue, are also endowed with a redshift determination.

\begin{figure}
\begin{center}
\includegraphics[scale=0.42]{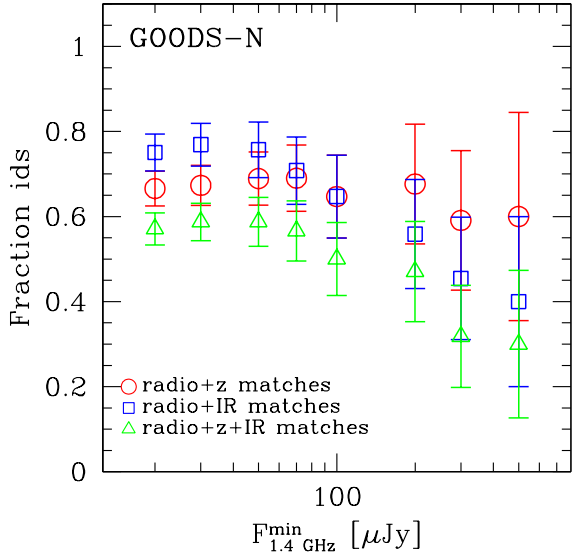}
\caption{Fraction of radio sources in the GOODS-N endowed with a redshift determination (red circles), with a FIR counterpart from the PEP catalogue (blue squares) and with both a FIR id and a redshift determination (green triangles) as a function of integrated 1.4 GHz flux. 
Error-bars correspond to Poissonian estimates.
\label{fig:fraction_GN}}
\end{center}
\end{figure}

As it was for GOODS-S, far infrared information for the GOODS-N radio sources were obtained through the PEP (Lutz et al. 2011) and  PEP/GOODS-H (Magnelli et al. 2013) catalogues. Once again, the maximum tolerance radius chosen to associate a radio source to a FIR counterpart was 4$^{\prime\prime}$. 
This procedure leads to 315, 1.4 GHz-selected, radio sources with a FIR counterpart either at 100 $\mu$m or at 160 $\mu$m. This corresponds to $\sim79$\% of the parent catalogue,  in good agreement with what found in the GOODS-South. 
Also in this case, the expected fraction of random associations  is less than 2\%.\\
The trend for the fraction of FIR identifications as a function of integrated radio flux is presented in Figure \ref{fig:fraction_GN} by the blue squares, while that for sources with a redshift determination by the red circles and that for sources with both a FIR counterpart and a redshift determination by the green triangles. In all cases errors correspond to Poissonian estimates. Note that, once again, while the fraction of sources with a redshift determination tends not to depend on flux at all radio fluxes, the fraction of radio sources with a FIR counterpart  stays constant, equal to the values 
of $\sim80$\%, only up to radio fluxes F$_{1.4\rm GHz}\sim 100$ $\mu$Jy. Beyond this value such a ratio starts to monotonically decline. The same trend is also  reflected in the behavior of the fraction of radio sources with both a redshift determination and a FIR counterpart.

Table~1 presents a summary of the properties of the 1.4 GHz-selected radio sources on all the three fields considered in this work.

\begin{table}
\begin{center}
\caption{Properties of the analysed samples. The various columns represent the three different fields considered in this work. The first row indicates the total number of radio sources selected at 1.4 GHz on regions of the sky covered by deep {\it Herschel}/PACS observations, N$_{\rm TOT}$, the second one the number of radio sources with either a spectroscopic or a photometric redshift determination, N$_{\rm z}$, the third row the number of radio sources with a FIR counterpart either at 100 $\mu$m or at 160 $\mu$m, 
N$_{\rm IR}$, and the last row the number of radio sources with both a FIR counterpart and a redshift determination, N$_{\rm IR+z}$.}
 \begin{tabular}{llll}
& Lockman Hole & GOODS-S & GOODS-N\\
\hline
\hline

N$_{\rm TOT}$&698&142&401\\
N$_{\rm z}$&402&114&267\\
N$_{\rm IR}$&422&115&315\\
N$_{\rm IR+z}$&260&87&229\\
\hline
\hline

\end{tabular}
\end{center}
\end{table}

\section{AGN selection via radio luminosity}
A tricky issue in the present work is the distinction between radio emission produced by the presence of an AGN  and that originating from star-formation processes. 


\begin{figure}
\begin{center}
\includegraphics[scale=0.42]{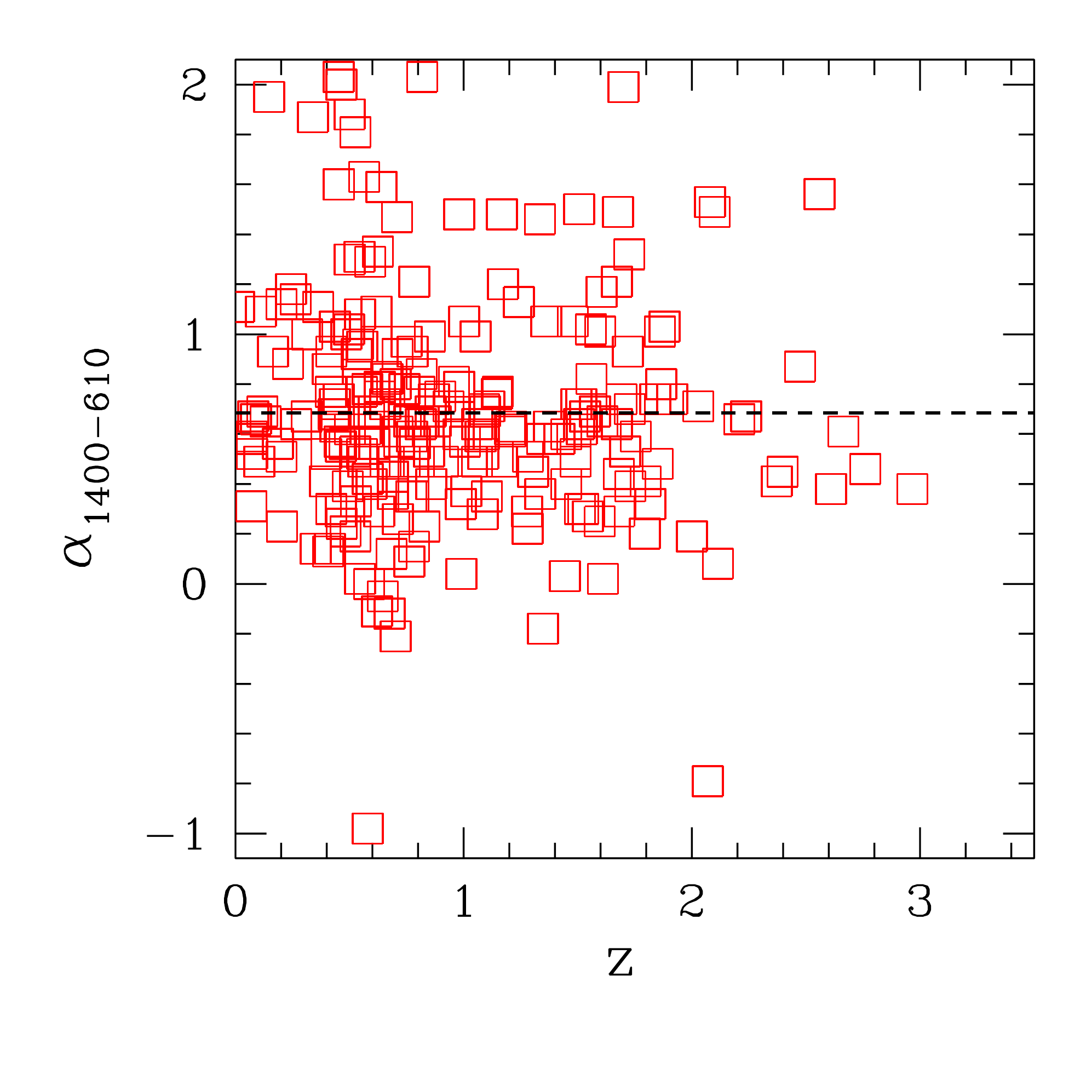}
\caption{Distribution of radio spectral indices between 1.4 GHz and 610 MHz as a function of redshift for the radio sources in the Lockman Hole Field with measured redshifts. The horizontal dashed line represents the average of the distribution 
$<\alpha>=0.685$.
\label{fig:alpha_LH}}
\end{center}
\end{figure}

The method we adopt here was already introduced in Magliocchetti et al. (2014) and is based on the  results of McAlpine, Jarvis \& Bonfield (2013) who used the optical and near infrared Spectral Energy Distributions of a sample of 942 radio sources  (out of 1054 objects selected at 1.4 GHz, with a completeness level of 91\%) from the VIDEO-XMM3 field to distinguish between star forming and AGN-powered galaxies and derive their redshifts. 
These authors provide luminosity functions for the two classes of sources up to redshifts $\sim 2.5$ and find different evolutionary behaviours, with star-forming objects evolving in luminosity in a much stronger way ($\propto (1+z)^{2.5})$ than radio-selected AGN ($\propto (1+z)^{1.2}$).
Investigations of their results show that the radio luminosity P$_{\rm cross}$ beyond which  AGN-powered galaxies become the dominant radio population scales with redshift roughly as
\begin{eqnarray}
\rm Log_{10}P_{\rm cross}(z)=\rm Log_{10}P_{0,\rm cross}+z,
\label{eq:P}
\end{eqnarray}
at least up to $z\sim 1.8$. $P_{0,\rm cross}=10^{21.7}$[W Hz$^{-1}$ sr$^{-1}$] is the value which is found in the local universe and which roughly coincides with the break in the radio luminosity function of star-forming galaxies (cfr Magliocchetti et al. 2002; Mauch \& Sadler 2007). Beyond this value,  their  luminosity function steeply declines, and the contribution of star-forming galaxies to the total radio population is drastically reduced to a negligible percentage. The same trend is true at higher redshifts, and since the radio luminosity function of star-forming galaxies drops off in a much steeper way than that of AGN at all $z$, we expect the chances of contamination in the two populations to be quite low.

\begin{figure*}
\includegraphics[scale=0.40]{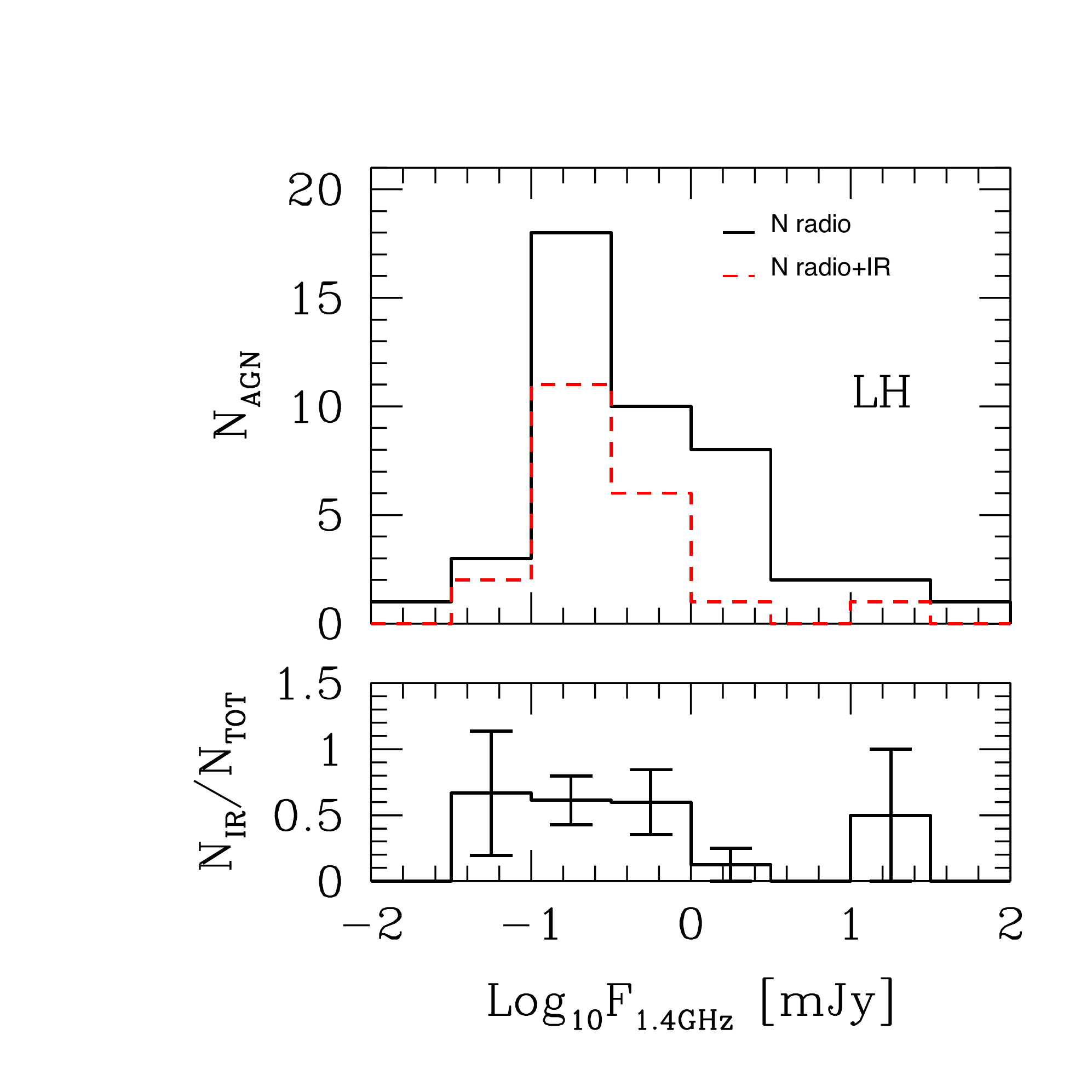}
\includegraphics[scale=0.40]{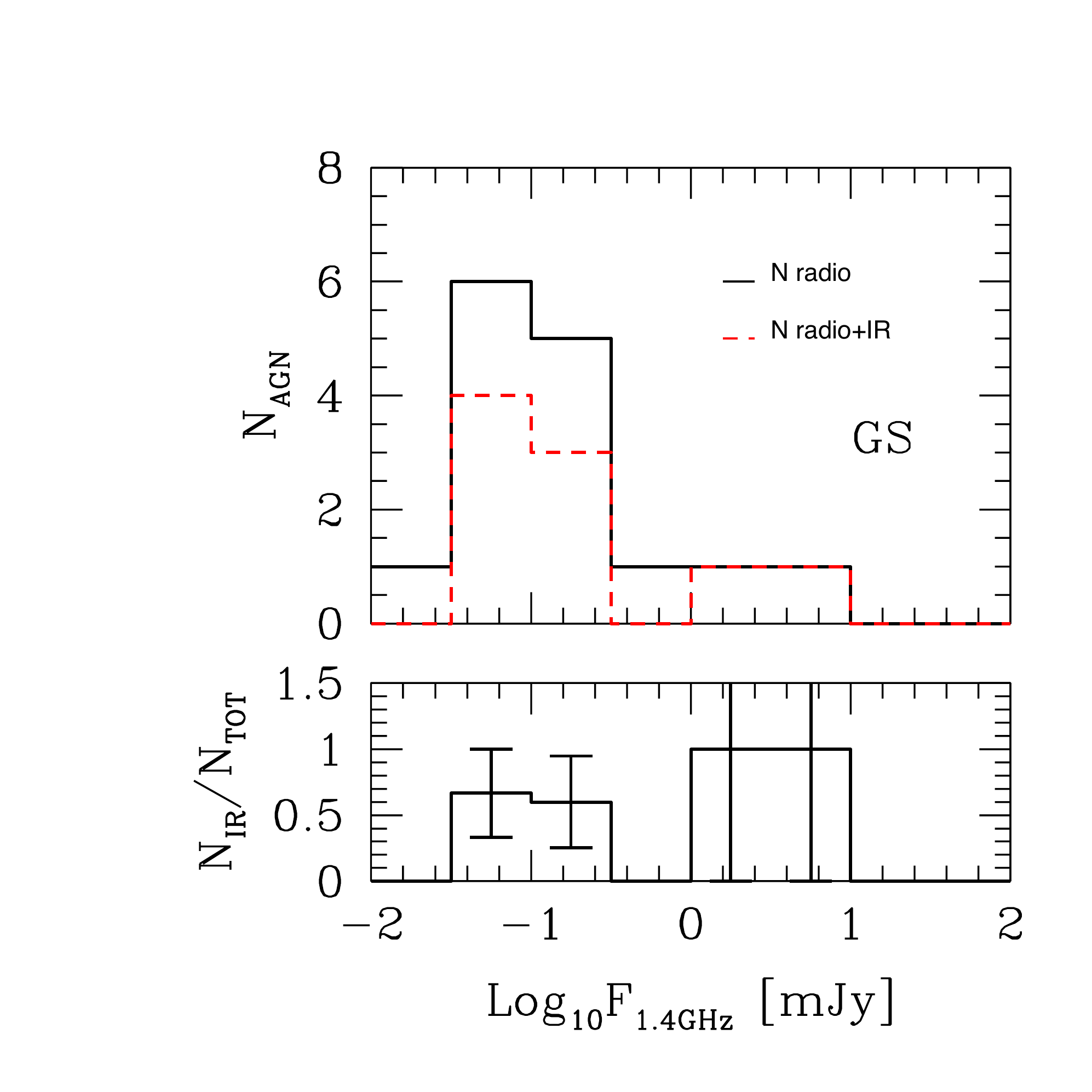}
\includegraphics[scale=0.40]{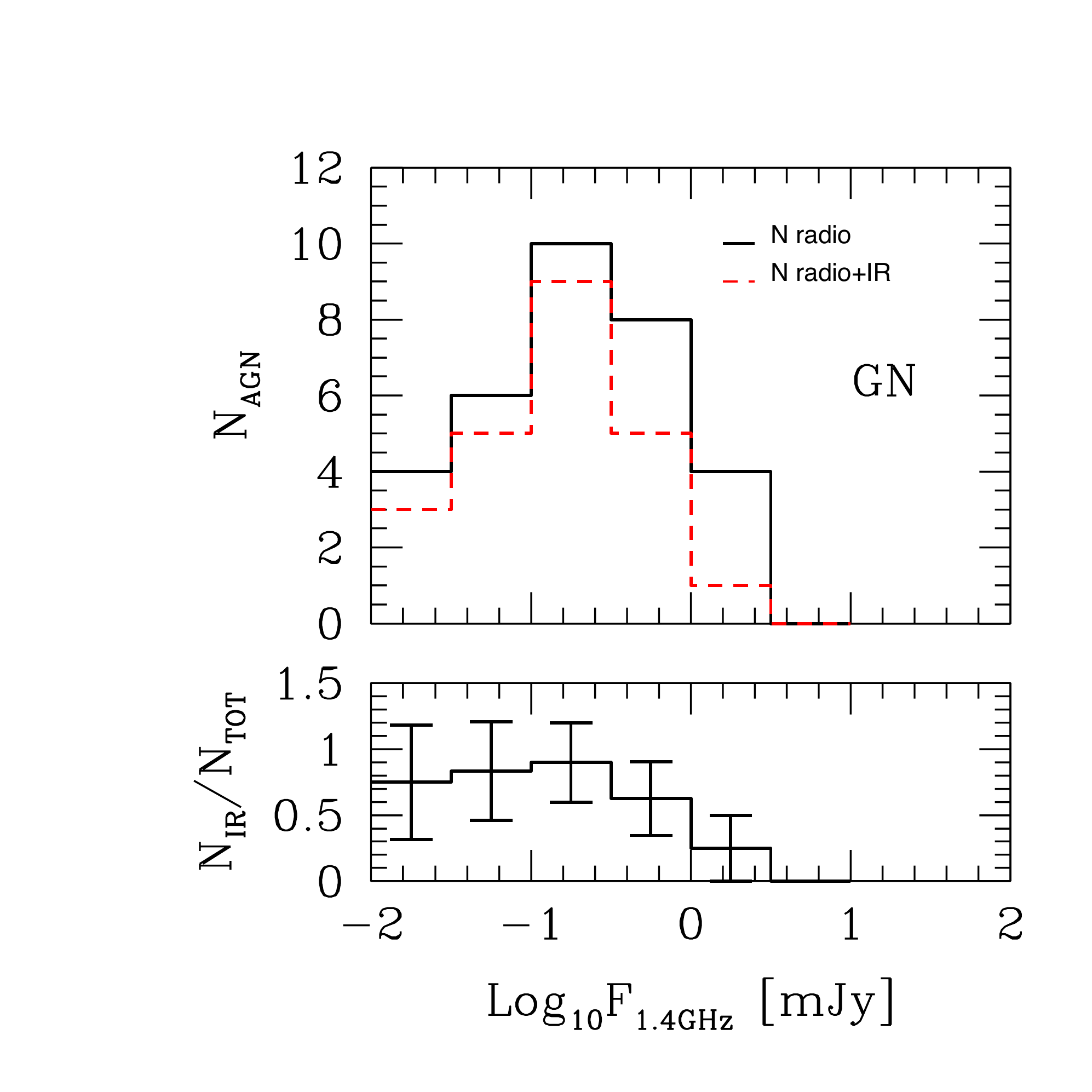}
\includegraphics[scale=0.40]{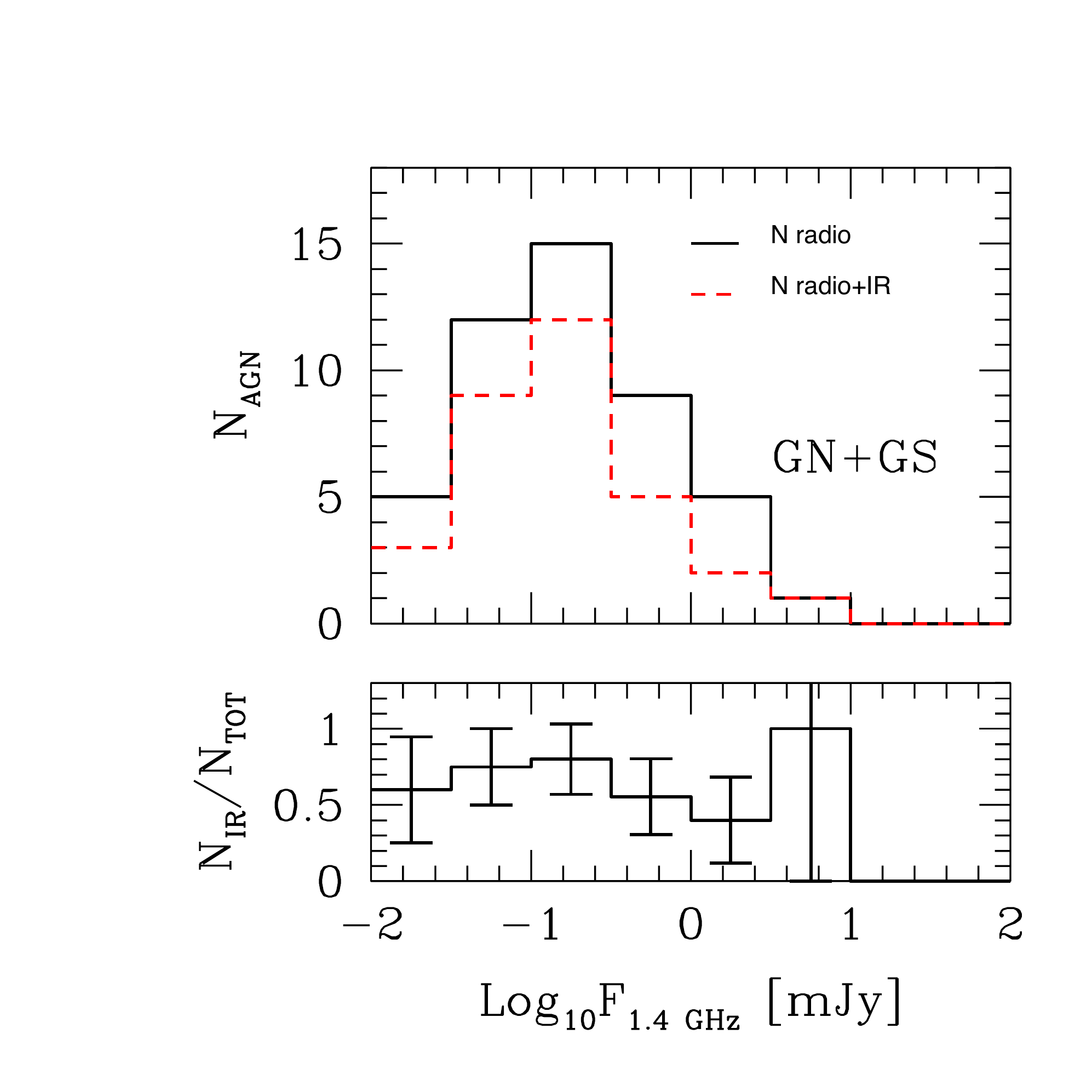}
\caption{Top panels: distribution of radio-emitting AGN as a function of 1.4 GHz flux. Clockwise, the panels show sources in the Lockman Hole (LH), in the GOODS-S (GS), in the combined GOODS-N and GOODS-S fields (GN+GS), and in the GOODS-N (GN). The solid black lines represent the whole population of radio-emitting AGN selected  on the basis of their radio luminosity, while red dashed lines show the subsets which also have counterparts at FIR wavelengths. Bottom panels: fraction of radio emitting AGN with a FIR counterpart as a function of radio flux. Error-bars correspond to 1$\sigma$ Poissonian estimates.
\label{fig:AGNvsflux}}
\end{figure*}

\begin{figure*}
\begin{center}
\includegraphics[scale=0.42]{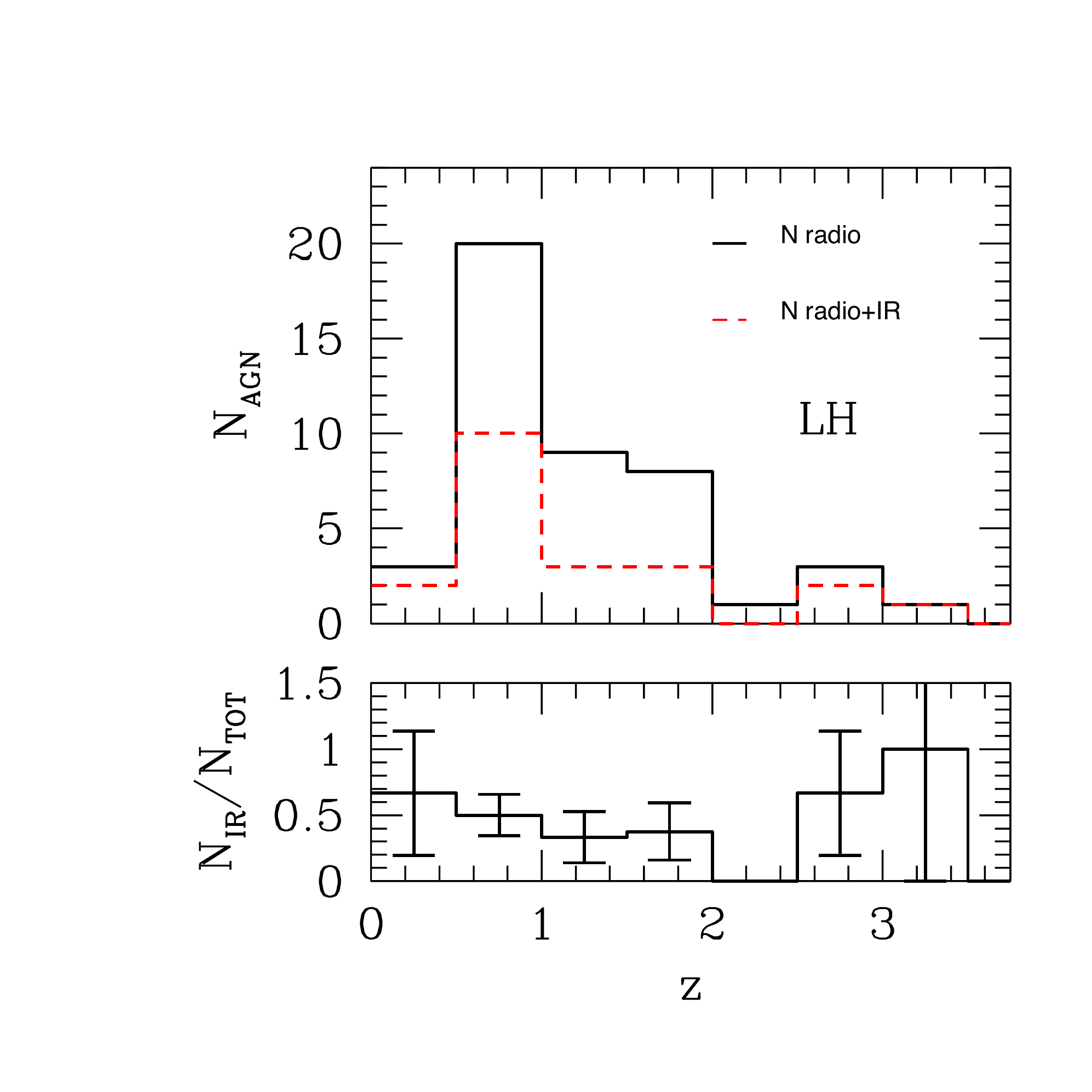}
\includegraphics[scale=0.42]{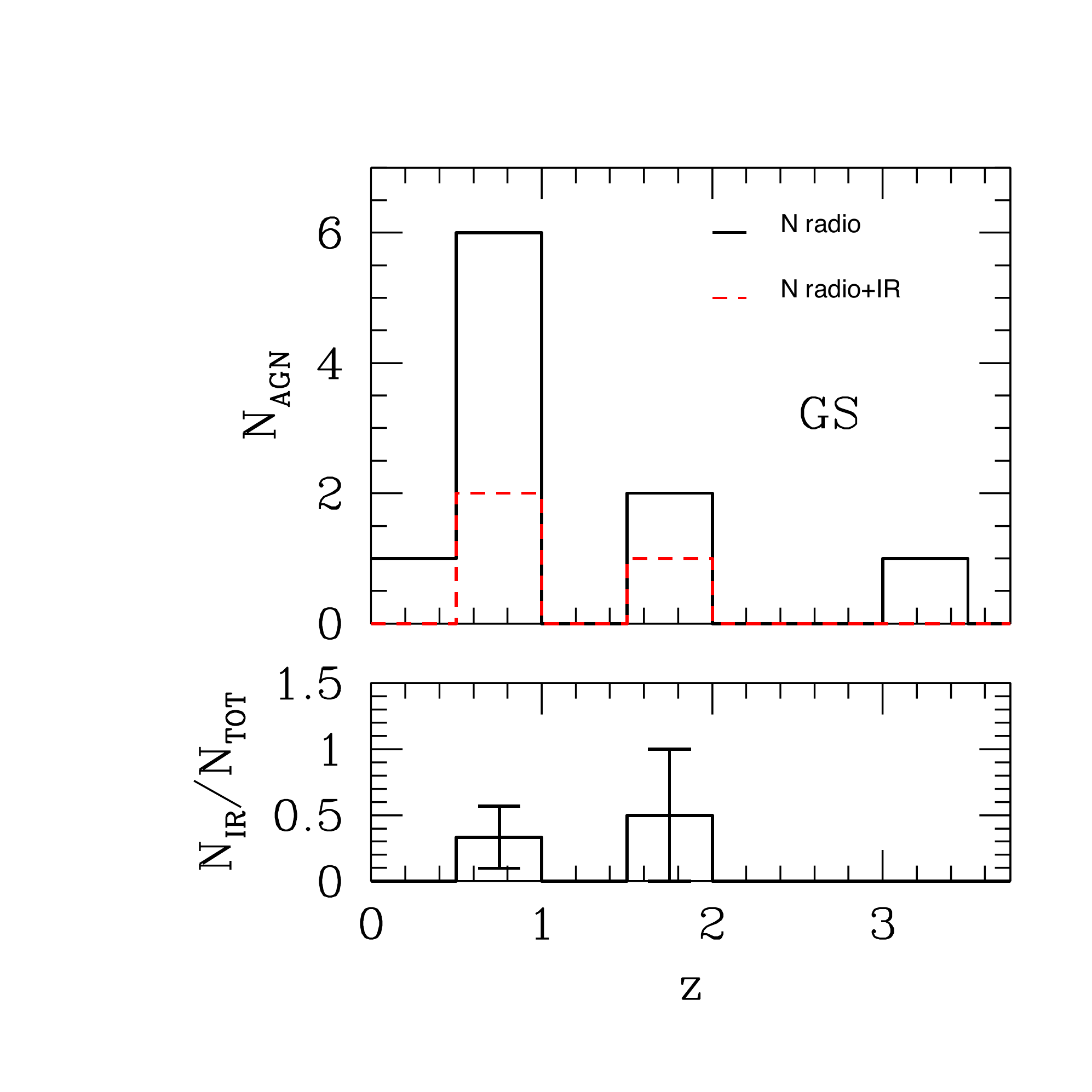}
\includegraphics[scale=0.42]{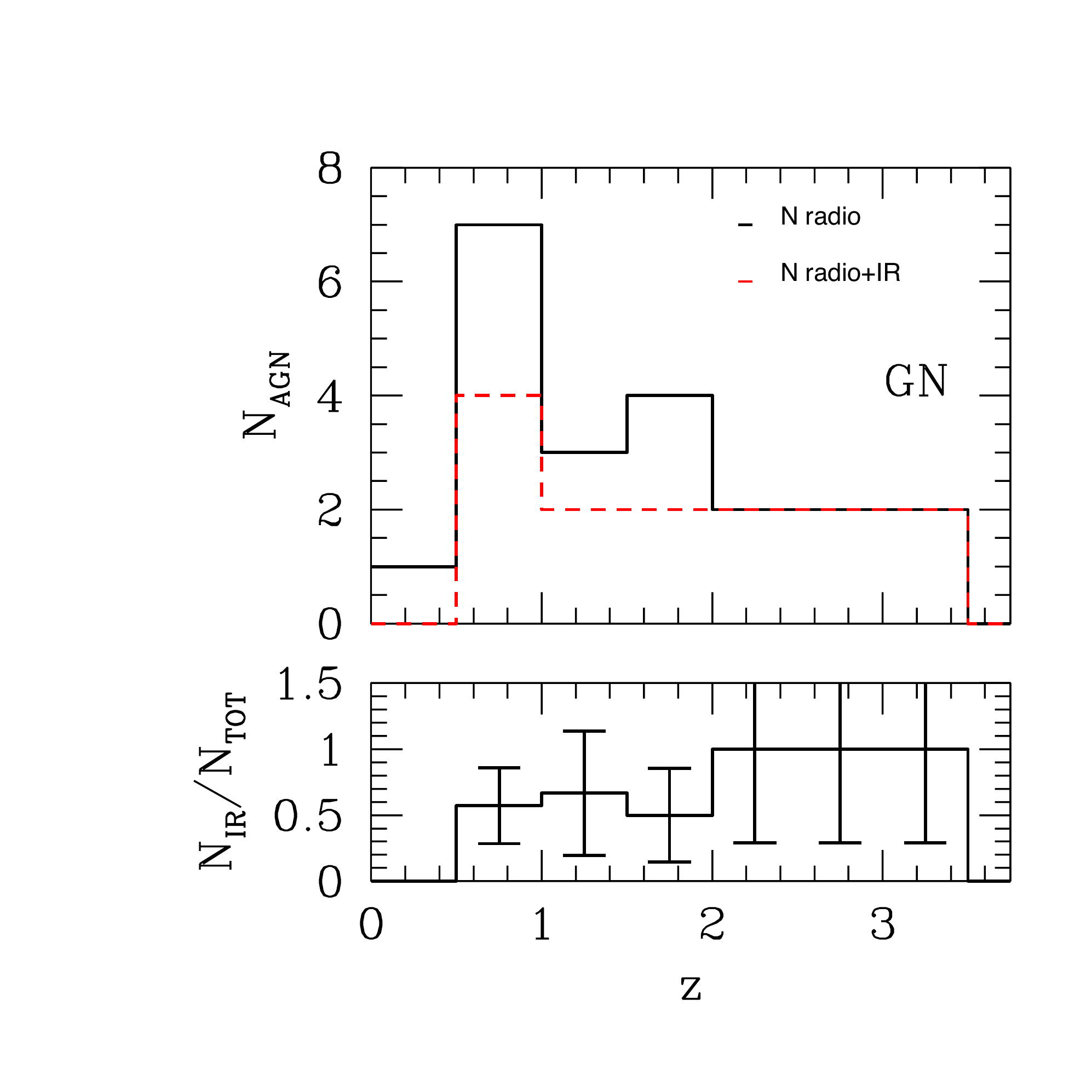}
\includegraphics[scale=0.42]{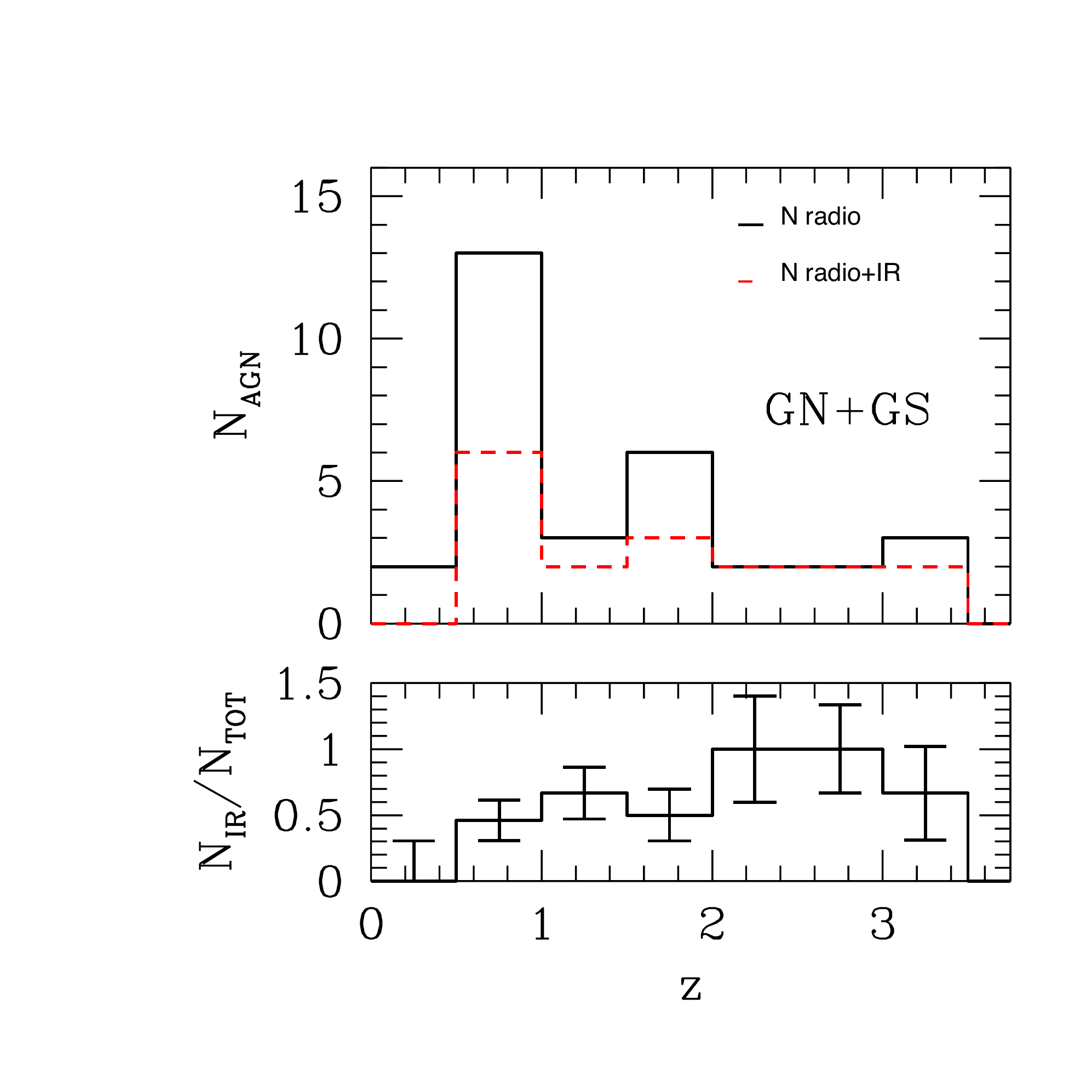}
\caption{Top panels: distribution of radio-emitting AGN as a function of redshift. Clockwise, the panels show sources in the Lockman Hole (LH), in the GOODS-S (GS), in the combined GOODS-N and GOODS-S fields (GN+GS), and in the GOODS-N (GN). The solid lines indicate the whole population of radio-emitting AGN selected on the basis of their radio luminosity, while red dashed lines show the subsets which also have counterparts at FIR wavelengths. Bottom panels: fraction of radio emitting AGN with a FIR counterpart as a function of redshift. Error-bars correspond to Poissonian estimates.
\label{fig:AGNvsz}}
\end{center}
\end{figure*}

Radio luminosities for our three samples of radio-selected sources endowed with a redshift estimate have been calculated according to the relation:
\begin{eqnarray}
\rm P_{1.4 \rm GHz}=\rm F_{1.4 \rm GHz} D^2 (1+z)^{3+\alpha},
\end{eqnarray}
where the result is in [W Hz$^{-1}$ sr$^{-1}$] units, D is the angular diameter distance and $\alpha$ is the spectral index of the radio emission ($\rm F(\nu)\propto \nu^{-\alpha}$). \\
As the Lockman Hole region has been imaged by Ibar et al. (2009) both at 1.4 GHz and at 610 MHz, most of the radio sources in this field have $\alpha$ estimates. 
The same is not true for either the GOODS-N or the GOODS-S fields. In these latter cases we then adopted the average value $\alpha=0.7$ found for similar surveys (e.g. Randall et al. 2012 and references therein) both for star-forming galaxies and for AGN emission. Such an assumption should not be too far from the truth as 1) radio sources in the GOODS samples are faint, therefore the chances of finding a large number of bright, flat spectrum AGN are rather thin and 2) recent results report values  $\alpha\simeq 0.7$ also for star-forming galaxies at $z\simeq 2$ (Ibar et al. 2010), similar to what found locally for the same population (Condon 1992). Furthermore, the value of $\alpha=0.7$ is in perfect agreement with the average value found for sources in the Lockman Hole ($<\alpha>=0.685$), independent of flux and redshift, as clearly indicated in Figure \ref{fig:alpha_LH}.

We then distinguished between AGN-powered galaxies and star-forming galaxies by means of equation (\ref{eq:P}) for $z\le 1.8$ and by fixing $\rm Log_{10}P_{\rm cross}(z)=23.5$ [W Hz$^{-1}$ sr$^{-1}$ ] at higher redshifts (cfr McAlpine, Jarvis \& Bonfield 2013). 
This procedure identifies 357 star-forming galaxies and 45 AGN (corresponding to 11\% of the total radio population) in the Lockman Hole, 89 star-forming galaxies and 15 AGN (corresponding to 13\% of the total radio population) in the GOODS-S,  235 SF and 32 AGN (corresponding to 12\% of the total radio population) in the GOODS-N.
Note that, due to the adopted selection criteria and thanks to the great depths of the considered surveys, {\it all} the AGN samples are {\it complete} with respect to radio selection at all probed redshifts, i.e. the considered samples  include  {\it all} radio-emitting AGN endowed with a redshift determination. 

The AGN distributions as a function of their 1.4 GHz fluxes are presented in the four top panels of Figure~\ref{fig:AGNvsflux} by the solid black lines, where the bottom-right plot refers to the combined GOODS-N and GOODS-S fields, added together because of the very similar depths of the radio, FIR and multi wavelength surveys. The red, dashed lines represent the distributions of radio-emitting AGN which also possess a FIR counterpart. The histograms in the four bottom panels show the ratio between the full radio-selected AGN population and the sub-class of FIR emitters. As it is possible to notice, in all cases but GOODS-S, where the clarity of the results is hampered by the paucity of sources, there is a systematic trend for the fractions of radio-selected AGN which are also FIR emitters to roughly stay constant at the lowest radio fluxes probed by the different surveys and then possibly decline above F$_{1.4 \rm GHz}\sim 0.5-1$~mJy in the GOODS  fields (statistical significance $\sim 2\sigma$) and F$_{1.4 \rm GHz}\sim 1$~mJy in the Lockman Hole (statistical significance above the 3$\sigma$ level). Note that, even if the general trend is the same in all fields,  the success rate for FIR identifications indeed varies from field to field, being lower ($\sim 50$\%) in the Lockman Hole where FIR observations are shallower, and higher ($\sim 80\%$) in the two GOODS fields which present the deepest FIR observations obtained so far.

The AGN distributions as a function of redshift  are presented in the four top panels of Figure~\ref{fig:AGNvsz} by the solid black lines, where again the bottom-right plot refers to the combined GOODS-N and GOODS-S fields. The red dashed lines represent the distributions of radio-emitting AGN which also possess a FIR counterpart. The histograms in the four bottom panels show the ratio between the full radio-selected AGN population and the sub-class of FIR emitters. As it is possible to notice, at variance with the former case, there is a clear indication for the fractions of radio-selected AGN which are also FIR emitters to remain constant with look back time, up to the highest redshifts and in all the fields probed by our analysis. Note that once again the success rate for FIR identifications varies from field to field, being lower ($\sim 50$\%) in the Lockman Hole and higher ($\sim 80\%$) in the two GOODS fields.

\begin{table}
\begin{center}
\caption{Properties of the analysed samples. The various columns represent the three different fields considered in this work. The first and second rows respectively indicate the total number of radio-emitting AGN, N$_{\rm AGN}$, and star-forming galaxies, N$_{\rm SF}$, selected on the basis of their radio luminosity. The numbers in brackets are for those sources which are also FIR emitters.} 
 \begin{tabular}{llll}
 \\
& Lockman Hole & GOODS-S & GOODS-N\\
\hline
\hline

N$_{\rm AGN}$&45 (21)&15 (8)&32 (23)\\
N$_{\rm SF}$&357 (238)&89 (79)&235 (206)\\
\hline
\hline

\end{tabular}
\end{center}
\end{table}

Some interesting pieces of information on the general properties of radio-emitting AGN can then be gathered through investigation of Figures \ref{fig:AGNvsflux} and \ref{fig:AGNvsz}. In all considered fields their redshift distributions, which we remind, due to the exquisite depths of the radio surveys considered in this work, refer to {\it complete AGN samples selected in radio luminosity}, present a prominent peak at around $z\sim 1$ and then a possible secondary bump for $z\sim 2$. This is in agreement with the results found by Magliocchetti et al. (2014) for the COSMOS field. Furthermore, the redshift distribution of  AGN which also present FIR emission closely follows that of their parent population. Once again this is true for all the considered fields, as well as it was in COSMOS (Magliocchetti et al. 2014). 
It follows that, in all cases, the fraction of FIR emitters stays constant throughout the whole redshift range up to the earliest epochs probed by our analysis ($z\sim 3-4$), i.e. {\it there is no preferred age for radio-emitting AGN to be associated with FIR emission}. Indeed, the only appreciable difference amongst the different fields is the value of such a fraction, which is lower and equal to approximately 50\% in the Lockman Hole, while reaches values of about 70-80\% in the two GOODS fields where, we remind, FIR observations are the deepest ones available to date. 


A point which is worth stressing is the extremely high number of radio-emitting AGN which are found to possess a counterpart in {\it Herschel} maps. This can be appreciated both in Table 2 and in Figures \ref{fig:AGNvsflux} and \ref{fig:AGNvsz}. In fact, as already mentioned, one goes from the $\sim 50$\% of the Lockman Hole, to the $\sim80$\% of the GOODS fields. These figures have to be compared with the $\sim 40$\% found in COSMOS  (Magliocchetti et al. 2014), where the different percentages are only due to the different depths of FIR observations on the different fields. 
Allowing for  the relatively large uncertainties due to the small areas covered, these results imply that the overwhelming majority ($66\pm 12$\%, percentage which rises to $\sim 72$\% in the GOODS-N) of radio-emitting AGN in the GOODS-S and GOODS-N are imaged at FIR wavelengths down to fluxes 0.6 mJy at 100 $\mu$m and 1.3 mJy at 160 $\mu$m. This is a rather surprising result, since -- at least locally -- AGN selected at radio wavelengths have always being associated with passive galaxies, with little or no ongoing stellar activity. 

Another remarkable point that emerges from our analysis is that deeper FIR observations  do not pick up any new population nor they do single out any different property of the considered population.  In fact, the only effect of increased depth is that of evenly increasing the number of sources which are also FIR emitters, independent of  look-back time, radio flux and (as it will be more clear later on)  many other physical parameters which characterize the sources under examination.

\section{Properties of the inner engine}

\begin{figure*}
\begin{center}
\includegraphics[scale=0.42]{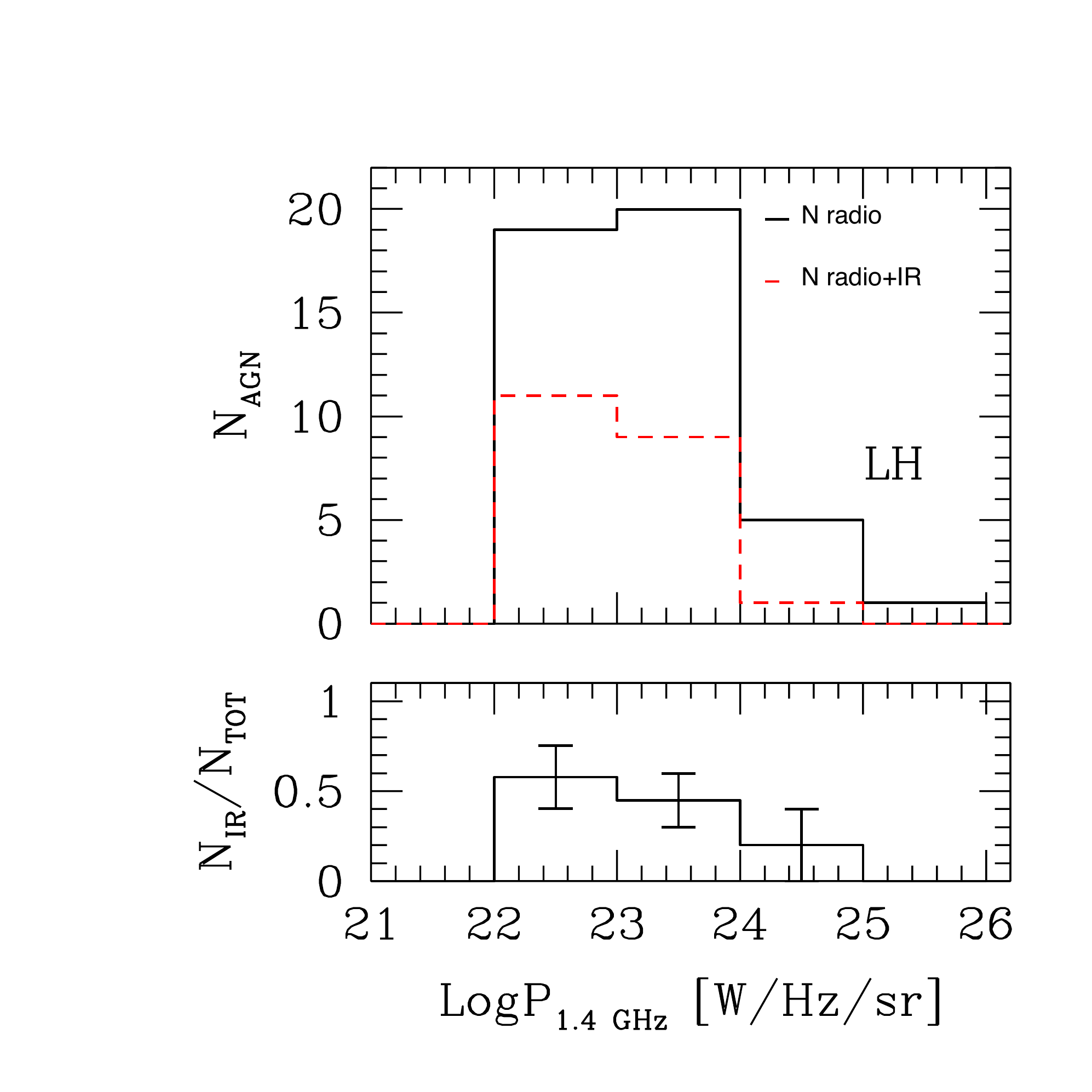}
\includegraphics[scale=0.42]{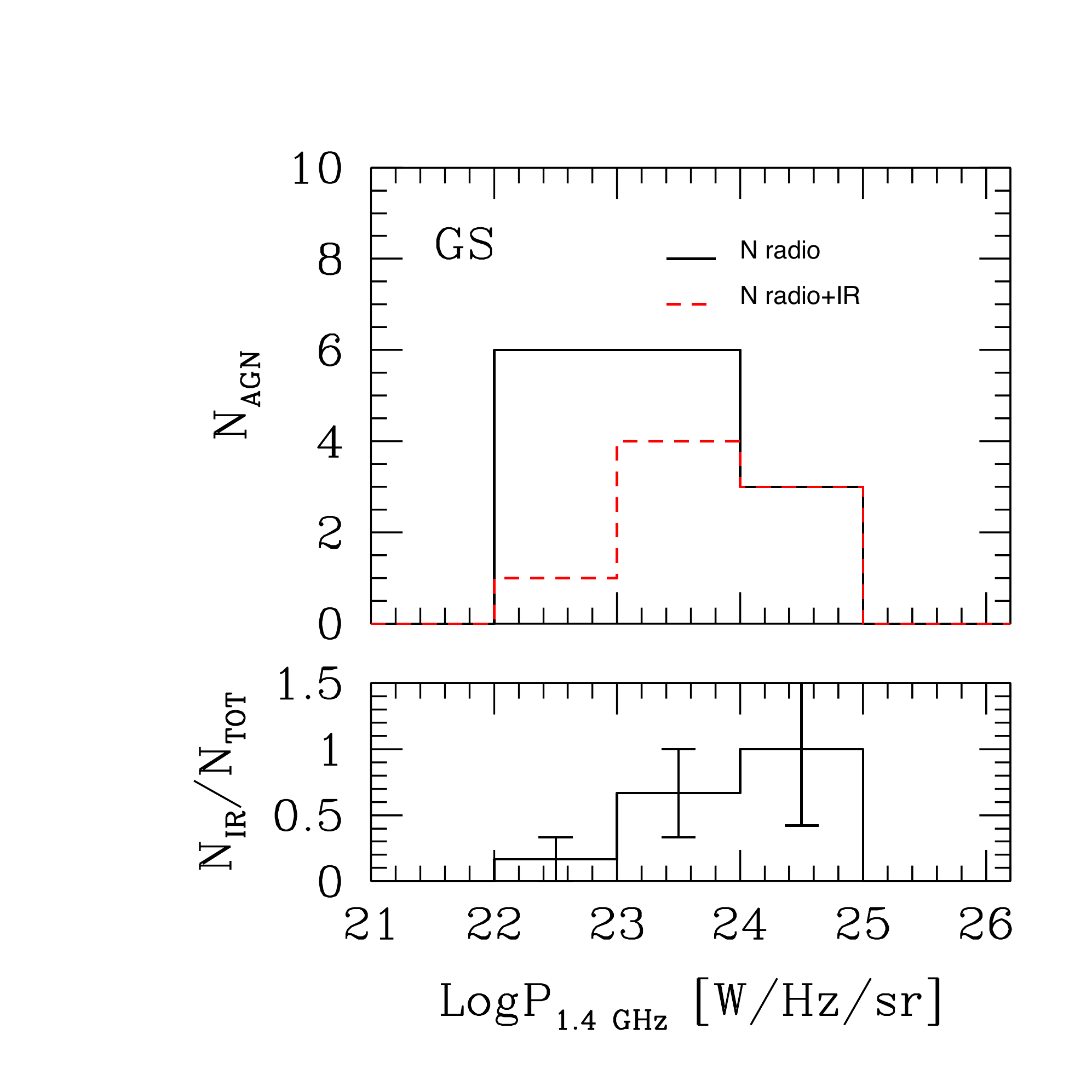}
\includegraphics[scale=0.42]{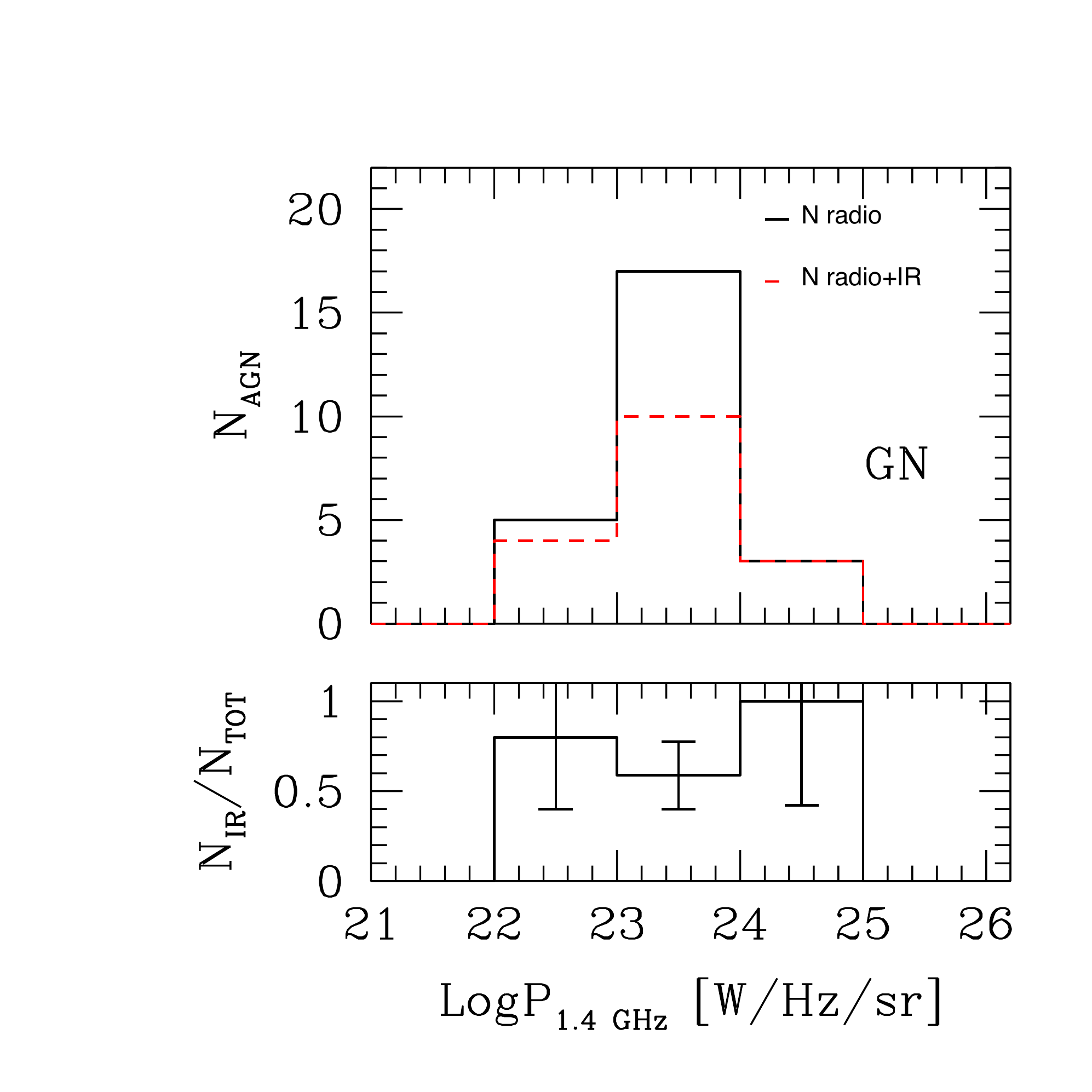}
\includegraphics[scale=0.42]{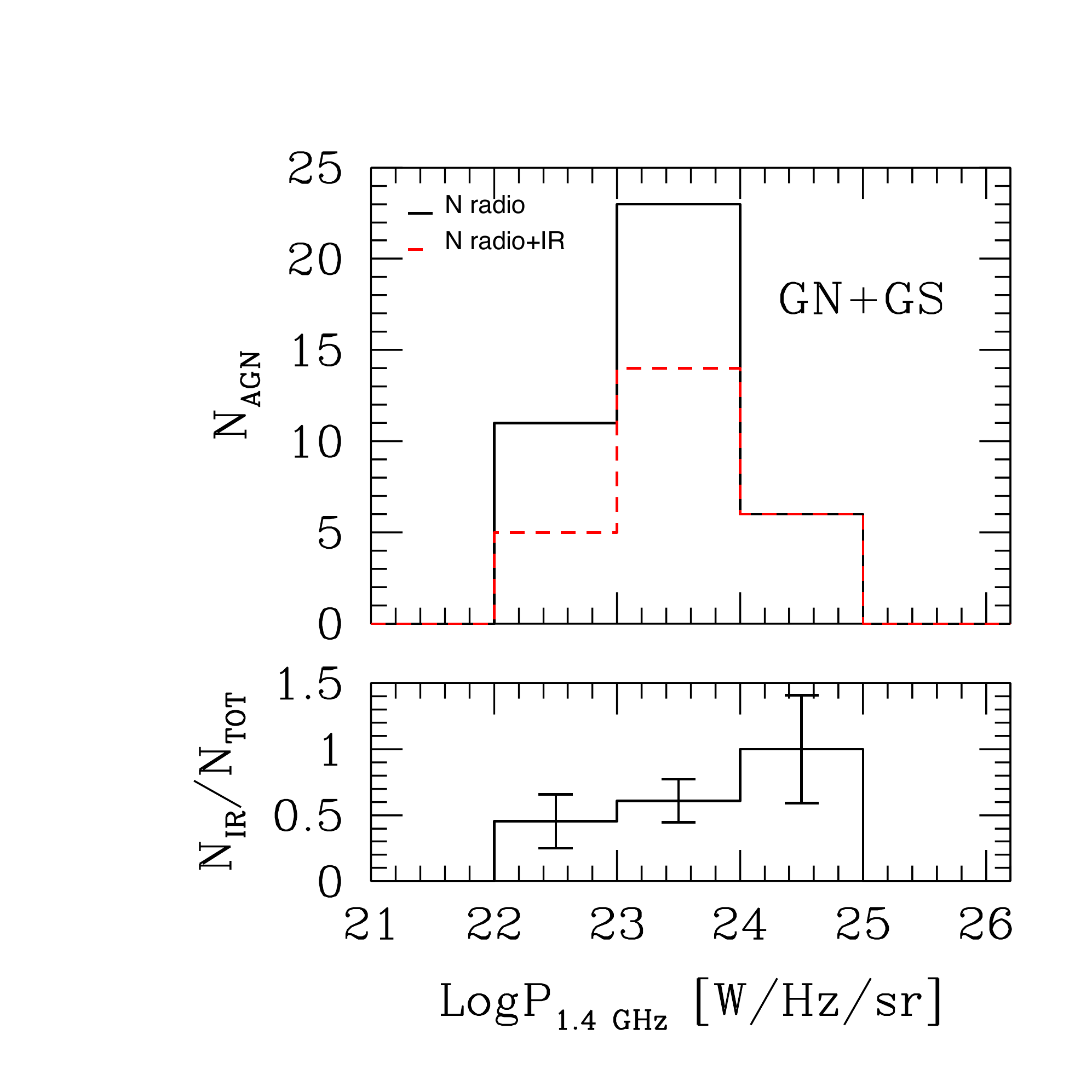}
\caption{Top panels: distribution of radio-emitting AGN as a function of of radio luminosity P$_{1.4 \rm GHz}$. Clockwise, the panels show sources in the Lockman Hole (LH), in the GOODS-S (GS), in the combined GOODS-N and GOODS-S fields (GN+GS), and in the GOODS-N (GN). The solid lines indicate the whole population of radio-emitting AGN selected on the basis of their radio luminosity, while red dashed lines show the subsets which also have counterparts at FIR wavelengths. Bottom panels: fraction of radio emitting AGN with a FIR counterpart. Error-bars correspond to Poissonian estimates.
\label{fig:AGNvsP}}
\end{center}
\end{figure*}

\begin{figure*}
\begin{center}
\includegraphics[scale=0.40]{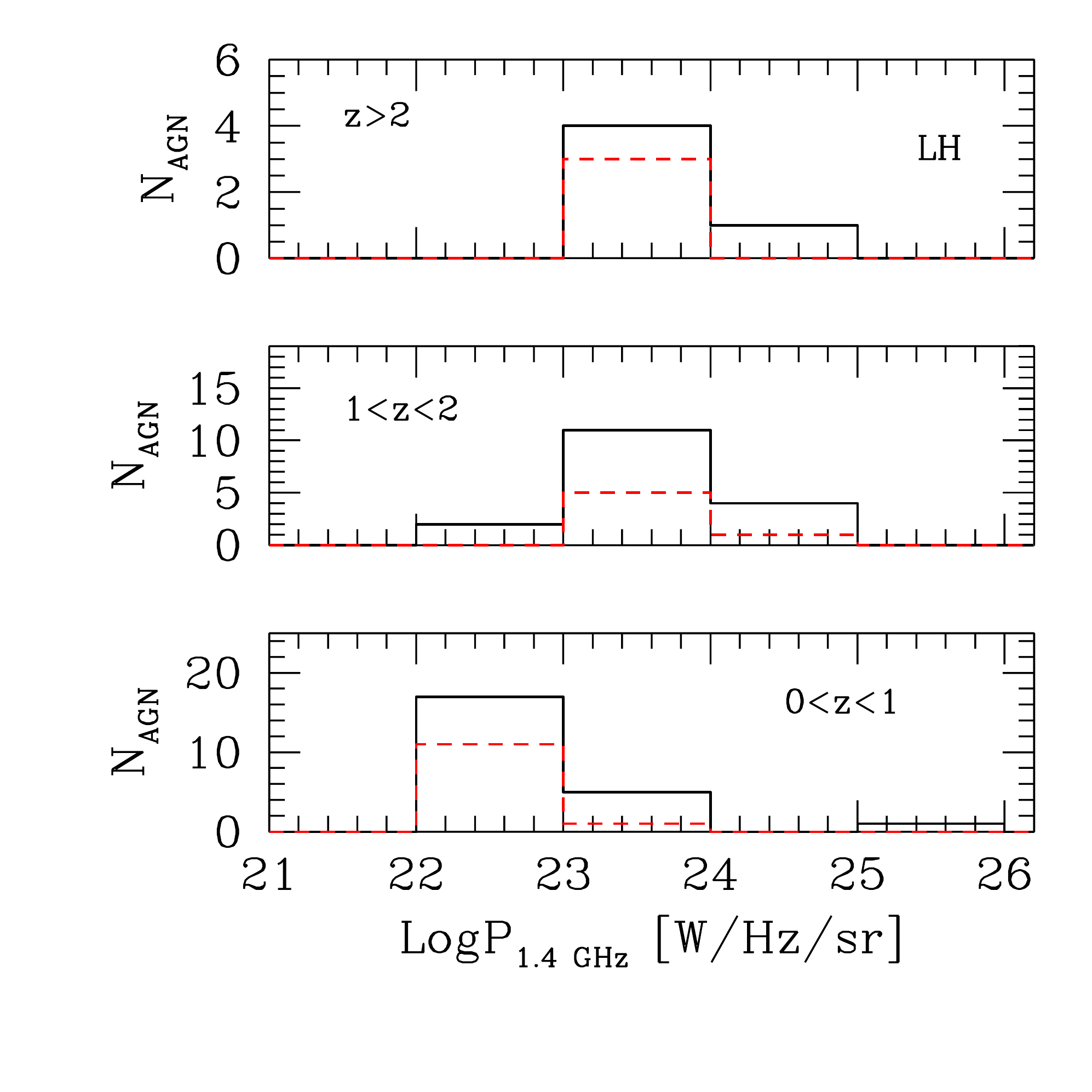}
\includegraphics[scale=0.40]{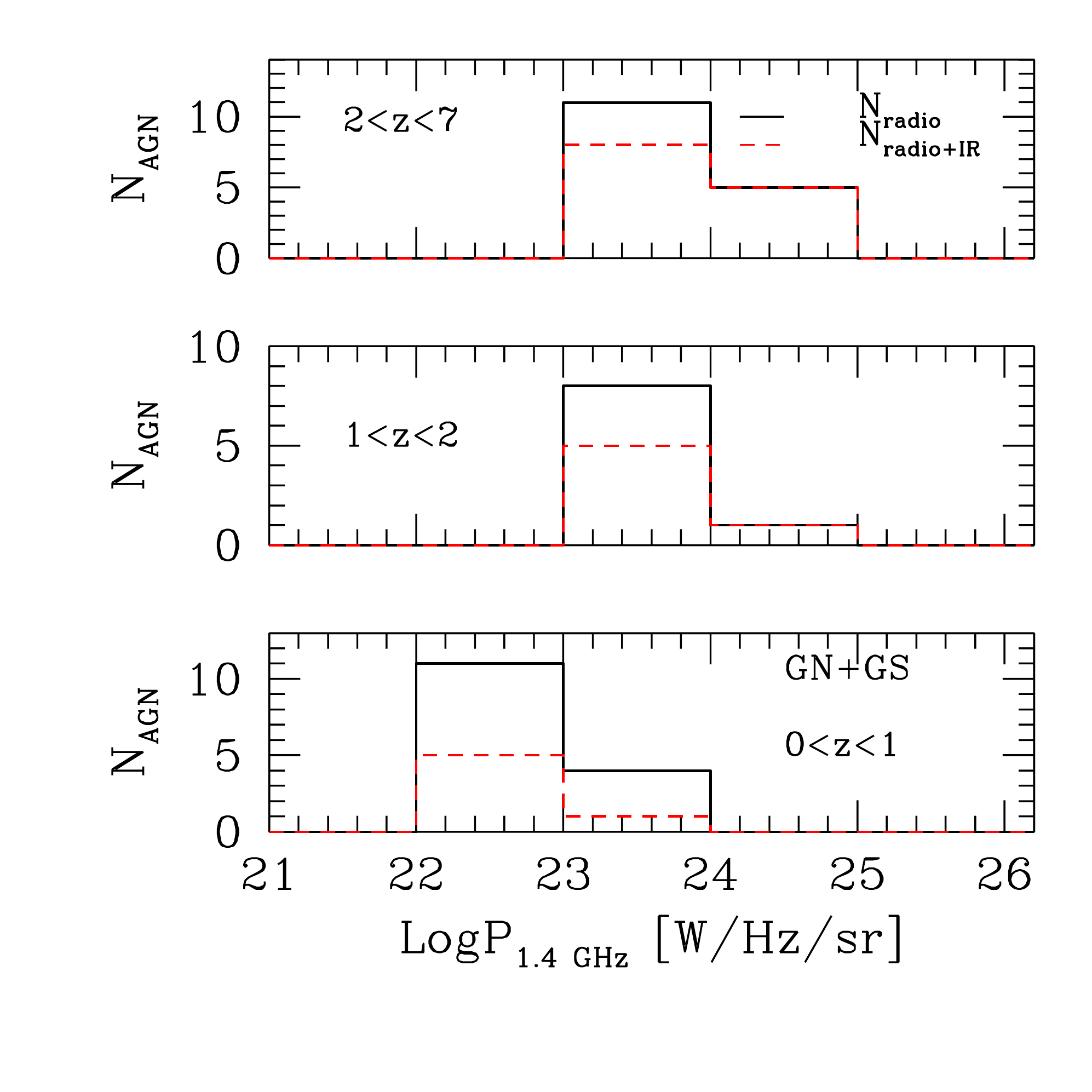}
\caption{Distribution of radio-emitting AGN as a function of  radio luminosity P$_{1.4 \rm GHz}$ in three different redshift bins. The left-hand panel shows sources in the Lockman Hole (LH), while that on the right-hand  side sources in the combined GOODS-N and GOODS-S fields (GN+GS). The solid lines indicate the whole population of radio-emitting AGN selected on the basis of their radio luminosity, while red dashed lines show the subsets which also have counterparts at FIR wavelengths.
\label{fig:Pdistrib_z}}
\end{center}
\end{figure*}

\begin{figure}
\begin{center}
\includegraphics[scale=0.40]{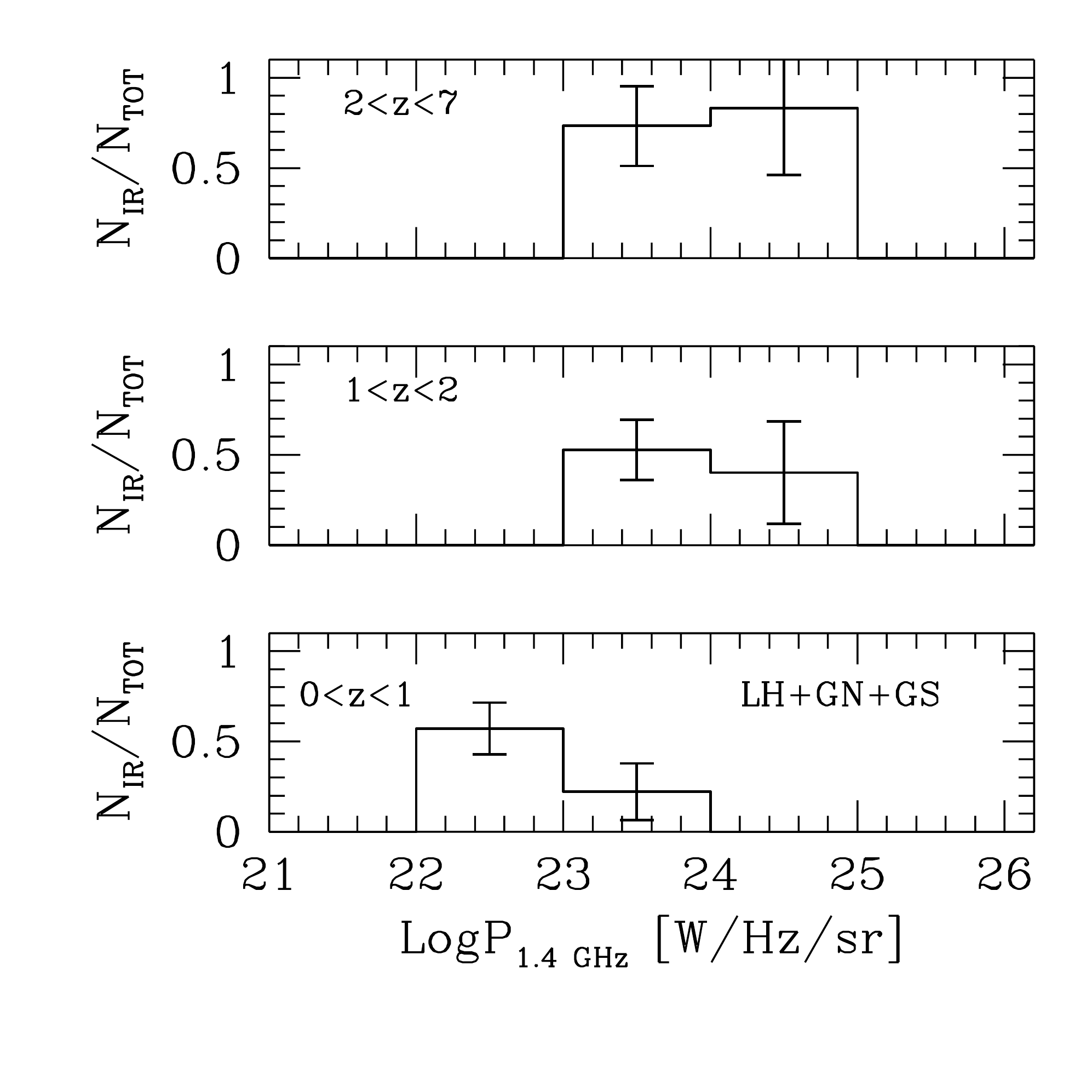}
\caption{Distribution of the fraction of FIR emitting AGN as a function of radio luminosity in three different redshift intervals. Error-bars correspond to Poissonian estimates.
 Data refer to all fields considered in this work.
\label{fig:AGNvsP_z}}
\end{center}
\end{figure}

Important information on the properties of the inner engine of radio-emitting AGN and assessment on whether there are any differences in the presence of FIR emission can be gathered by investigation of the distributions of radio luminosities and radio spectral indices for radio-emitting AGN both in the presence and in absence of FIR emission.

The top panels of Figure \ref{fig:AGNvsP} illustrate the distribution of AGN as a function of radio luminosity in the four fields considered here (Lockman Hole, GOODS-S, GOODS-N and the combined GOODS-S and GOODS-N). Once again, the solid lines represent the full population of radio-emitting AGN, while the dashed ones show the subsets which also have counterparts at FIR wavelengths. The bottom panels instead present the distribution  of the fraction of FIR emitters as a function of radio luminosity. 

In all the considered fields, there is a systematic drop of sources beyond luminosities P$_{1.4 \rm GHz}\simgt 10^{24}$ W Hz$^{-1}$ sr$^{-1}$. This is due to the fact that very bright radio-selected AGN are extremely rare, and finding them in small fields like those considered here is rather unlikely. 
Once again, the relative fraction of FIR emitters varies according to the depth of FIR observations on the field, and goes from the $\sim$ 50\% of the Lockman Hole to the $\simgt 70$ \% of the GOODS-N.

Some interesting information on the behavior of such sources can be found when splitting the samples into different redshift bins. This is done in Figure \ref{fig:Pdistrib_z} for both the Lockman Hole and the combined GOODS field. 
Despite the relative small number of sources, one can clearly see that the number of P$_{1.4 \rm GHz}\simgt 10^{23}$ W Hz$^{-1}$ sr$^{-1}$ radio-emitting AGN with a FIR counterpart  steadily increases as one moves back to the earlier universe. In more detail, for the combined GOODS fields we find that {\it all sources} with P$_{1.4 \rm GHz}\simgt 10^{24}$ W Hz$^{-1}$ sr$^{-1}$ and $z\simgt 1$ have a counterpart at FIR wavelengths and, at the same redshifts, between 70\% and 80\% of them are imaged in {\it Herschel} maps in the luminosity range $10^{23}$ W Hz$^{-1}$ sr$^{-1}\simlt$ P$_{1.4 \rm GHz}\simlt 10^{24}$ W Hz$^{-1}$ sr$^{-1}$. The case for the Lockman hole is substantially similar, even though the fraction of FIR emitters is not as high as in the GOODS fields as in this latter case FIR observations are shallower.

A better way to visualize the above trend is provided by Figure \ref{fig:AGNvsP_z} which shows the fraction of FIR emitting AGN as a function of radio luminosity in the same three different redshift bins as  in Figure \ref{fig:Pdistrib_z}. To increase the statistical significance of the results, we have plotted this quantity for all the three fields combined together. In this case, one can appreciate that  for redshifts greater than $\sim 1$ there is a substantial constancy with radio luminosity of the fraction of AGN which are also FIR emitters. In other words, {\it radio-selected AGN at redshifts $\simgt 1$ have all the same chances of  being FIR emitters independent of their radio luminosity}. However, this is not true at lower redshifts. In fact the plot shows that in the more local universe there is a ($\sim 2\sigma$) trend for FIR emitters to be mostly associated to low-luminosity sources. Such a probability then  decreases for brighter AGN. We further stress that, despite the different depth of FIR observations, the same behaviors are observed both in the Lockman Hole and in the combined GOODS fields, therefore strengthening the significance of our statement. 

\begin{figure*}
\begin{center}
\includegraphics[scale=0.4]{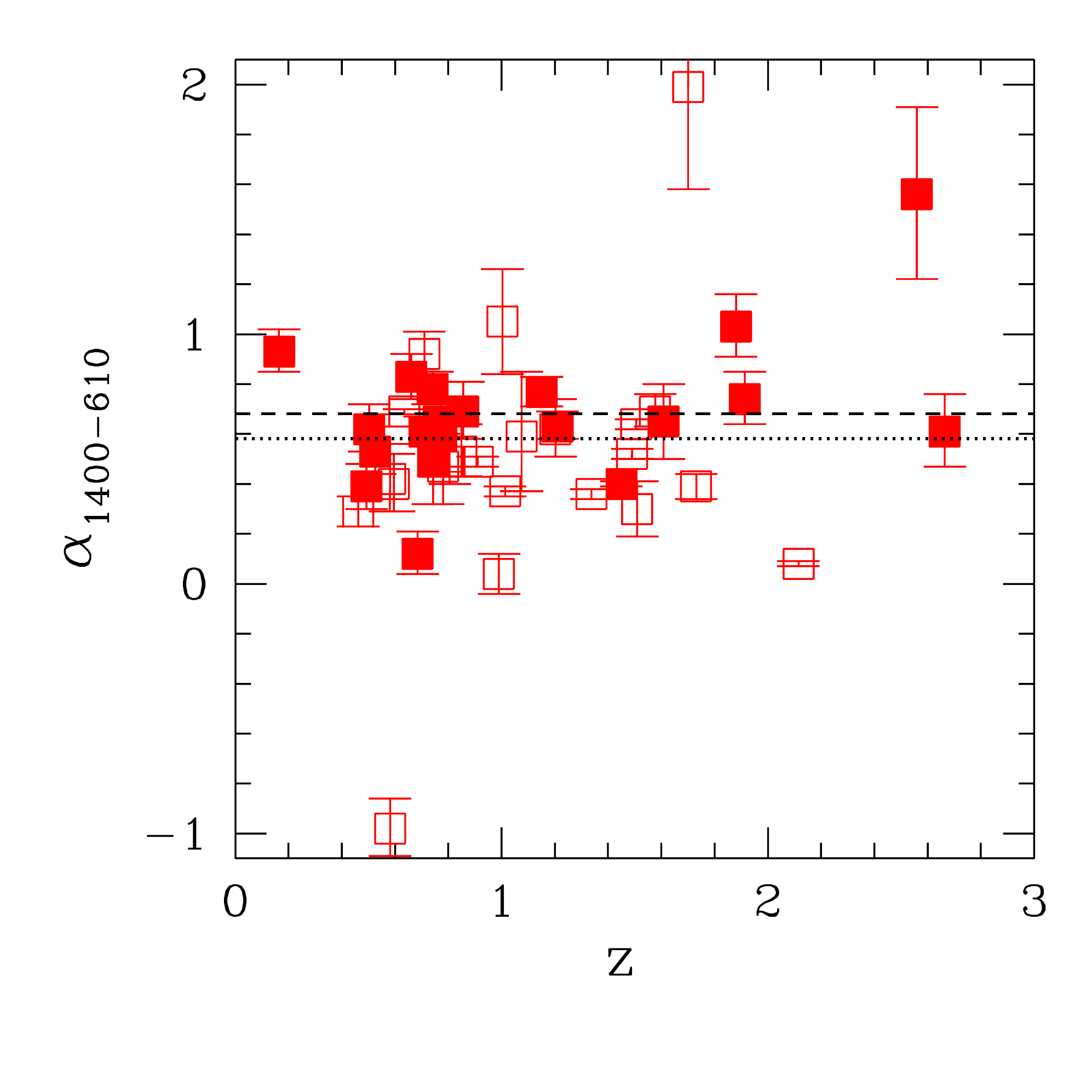}
\includegraphics[scale=0.4]{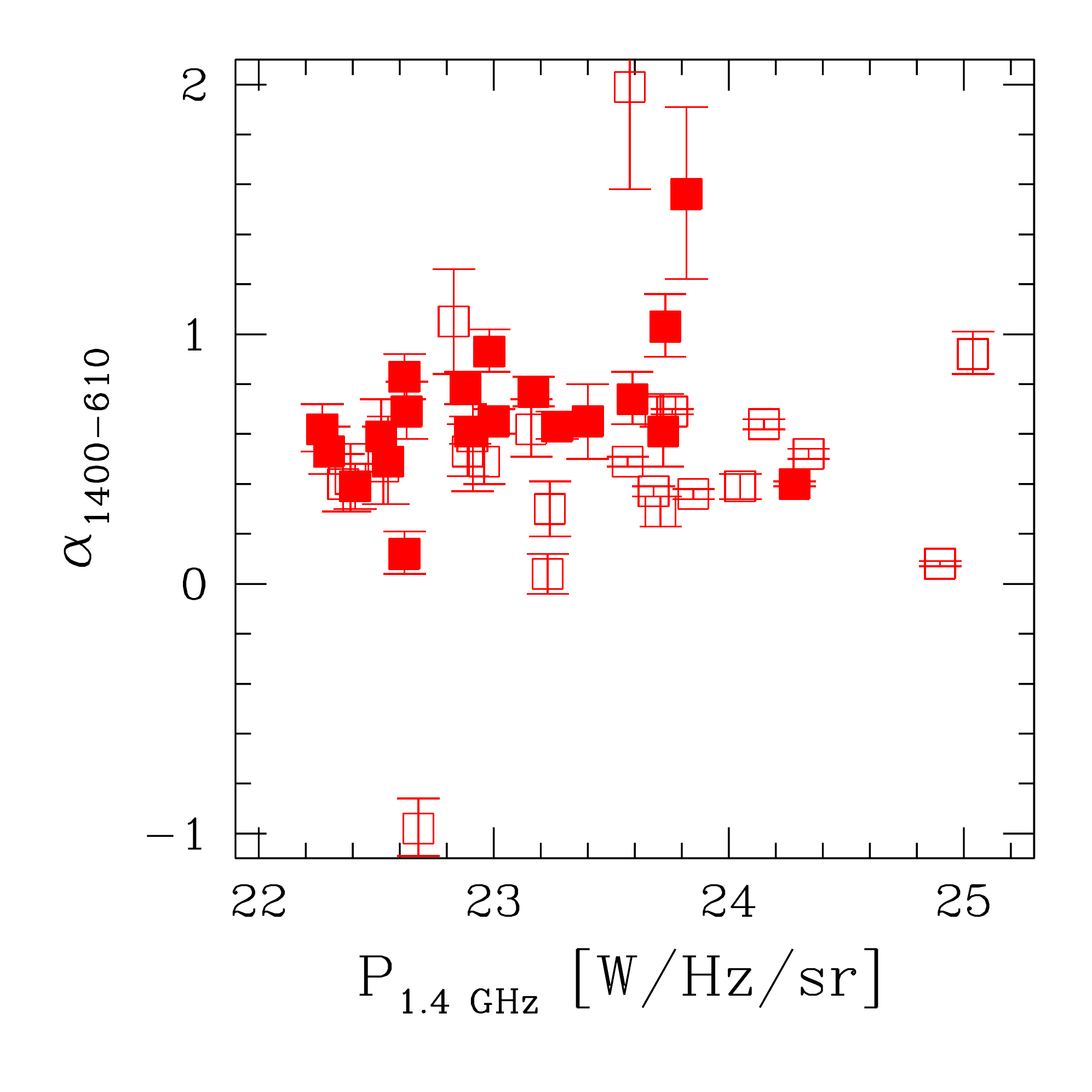}
\caption{Radio spectral indices between 1.4 GHz and 600 MHz for the population of radio-emitting AGN  as a function of redshift (left-hand plot) and radio luminosity (right-hand plot). The filled squares indicate the sub-population of FIR emitters. In the lefthand plot, the dotted line shows the average value of $\alpha_{1400-610}\simeq 0.58$ obtained for the whole population, the dashed one that of $\alpha_{1400-610}\simeq 0.68$ derived for FIR emitters. Data refer to the Lockman Hole.
\label{fig:alpha_AGN}}
\end{center}
\end{figure*}


What one can conclude from this analysis is that chances for an AGN which is active at radio wavelengths to also emit at FIR wavelengths is maximum at high ($z\simgt 1$) redshifts. For deep enough FIR observations such as those performed on the GOODS fields, this probability reaches the value of $\sim$ 100\%, independent of radio luminosity. This implies that {\it almost all} radio-selected AGN at $z\simgt 1$ are also the site of  FIR emission. The situation changes in the more local, $z\simlt 1$, universe, as in this latter case one finds that the relative fraction of FIR-active AGN  is lower and, most importantly, decreases as a function of radio luminosity. In other words, one has that in the local universe, mostly low-luminosity AGN are found to be associated with FIR emission. Only very few higher-power,  P$_{1.4 \rm GHz}\simgt 10^{23}$ W Hz$^{-1}$ sr$^{-1}$, sources instead appear on FIR maps. We note that this conclusions, although based on a relatively small sample of sources agree and further extend to fainter FIR emission those obtained by Magliocchetti et al. (2014) for the COSMOS field which, although covering a wider area, suffered of the limits due to relatively shallow FIR observations. They also agree with the local results which find radio-emitting AGN to reside in early-type galaxies, with little or no on-site ongoing star-forming activity. Our data however indicate that this is only true in the local universe, while in the more distant one FIR emission is extremely common amongst radio-active AGN. 

\begin{figure*}
\begin{center}
\includegraphics[scale=0.4]{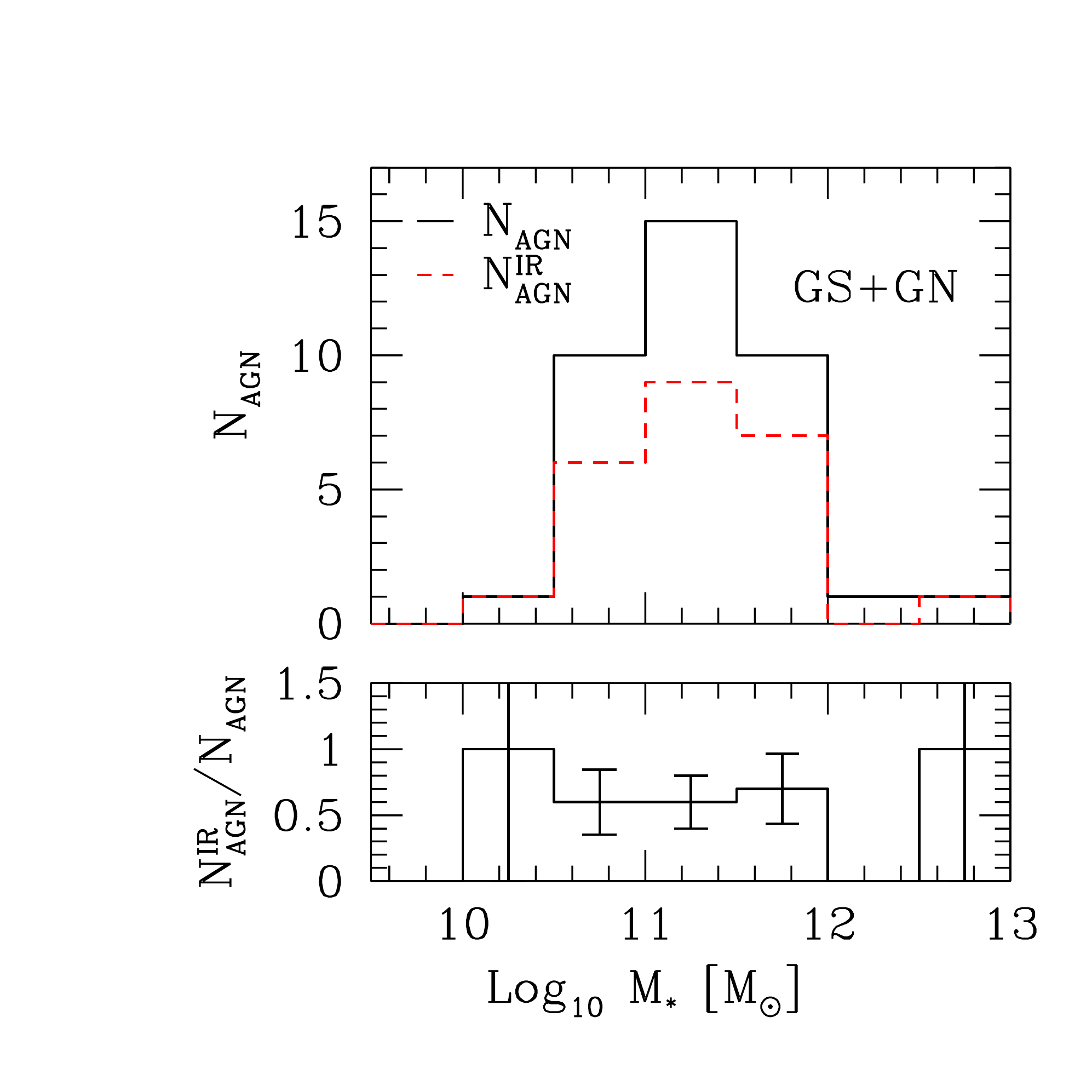}
\includegraphics[scale=0.35]{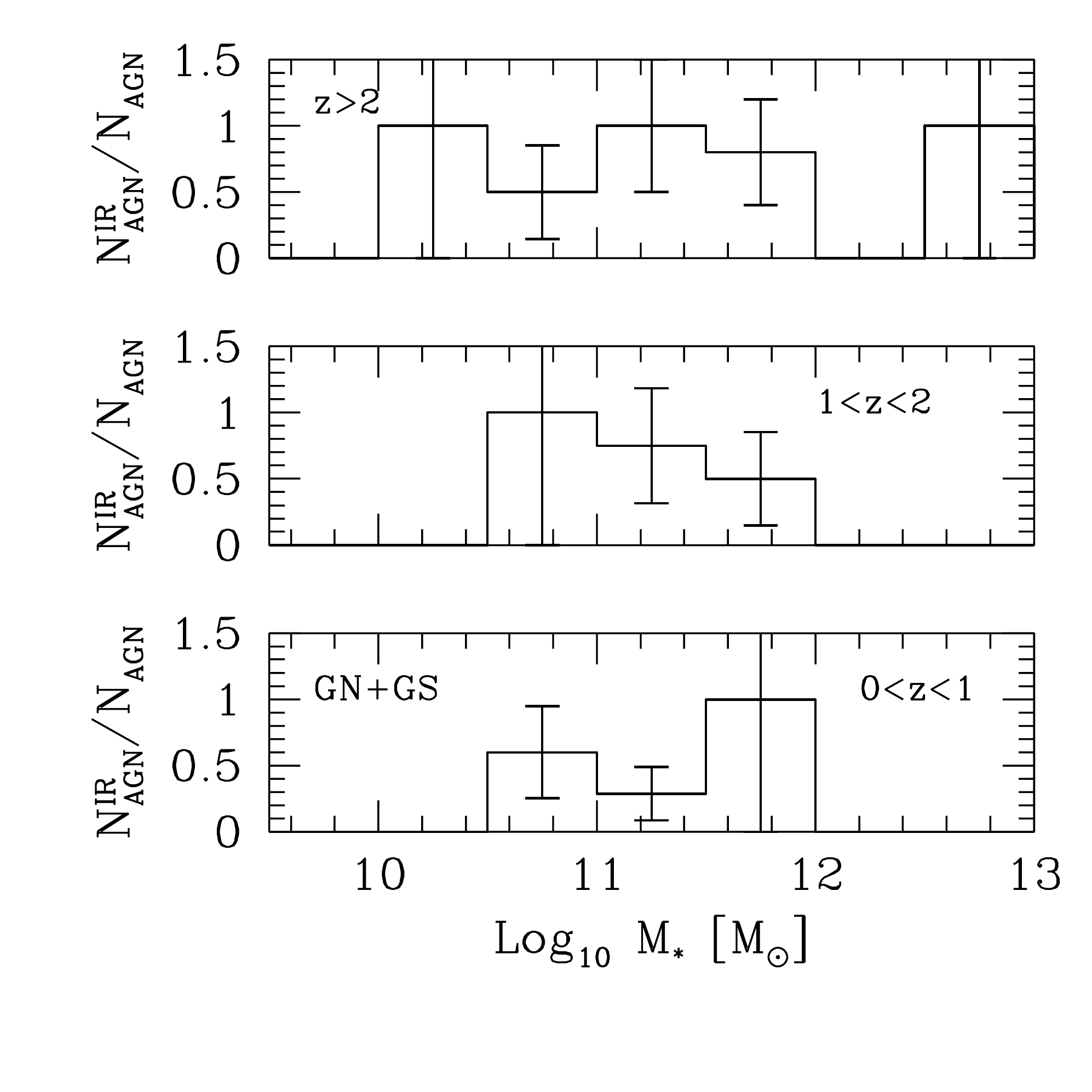}
\caption{Left-hand panel: stellar mass distribution for the galaxies hosts of radio-active AGN activity. The solid line shows  the whole galaxy population, while the dashed one represents the sub-class of FIR emitters. The bottom panel illustrates the ratio between these two quantities. Right-hand panel: fraction of radio-active galaxies which are also FIR emitters as a function of stellar mass in three different redshift bins. In all cases error-bars correspond to Poissonian estimates. 
Data refer to the combined GOODS-N and GOODS-S fields.
\label{fig:mass_GS}}
\end{center}
\end{figure*}

Further information on the engine of radio-active AGN can be obtained by investigating the spectral indices of their radio emission. As already discussed in \S 2.1, the Lockman Hole has been observed by Ibar et al. (2009) both at 1400 and at 600 MHz. 
Most of the observed sources (41 out of 45) are endowed with solid estimates of the quantity $\alpha_{1400-600}$. The trends for this quantity as a function of both redshift and radio luminosity are shown in Figure \ref{fig:alpha_AGN}. Squares represent the entire population, filled ones are for the sub-set of FIR emitters. The horizontal dotted  line shows the average $<\alpha>=0.58\pm 0.42$ value obtained for the entire population,  the dashed one represents  that of $<\alpha>=0.68\pm 0.29$ derived for FIR emitters. As it is clear also from these visual representations, there is no specific trend in the distribution of $\alpha$ values due to the eventual presence of FIR emission within the galaxies host of a radio-active AGN. 

All the distributions look identical, irrespective of the fact that the radio-active AGN also show emission at FIR wavelengths. This strongly suggests that, at least at radio wavelengths, the central engine responsible for the AGN phenomenon is oblivious to the FIR activity which is ongoing within the same host galaxy or, in other words, that FIR processes do not influence the activity of the central engine, at least that detected at radio wavelengths.

\section{Properties of the host galaxy}


\begin{figure}
\begin{center}
\includegraphics[scale=0.42]{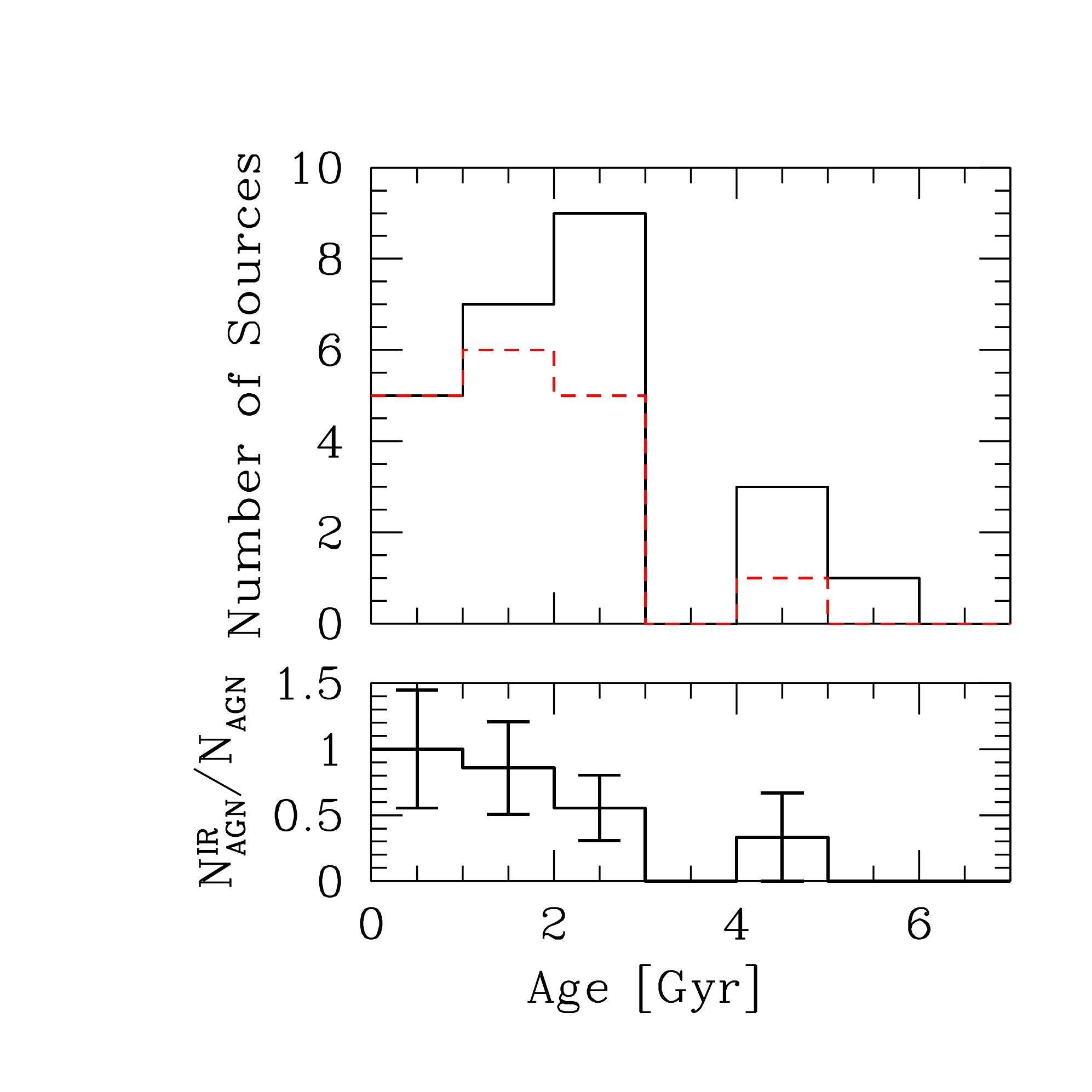}
\caption{Distribution of ages for galaxies in the GOODS-N hosting a radio-active AGN. The solid histogram in the top panel shows the trend for the entire AGN population, while the dashed one that for the sub-set of FIR emitters. The bottom panel presents the ratio between these two  quantities.  Error-bars correspond to Poissonian estimates.
\label{fig:age}}
\end{center}
\end{figure}

\begin{figure}
\begin{center}
\includegraphics[scale=0.42]{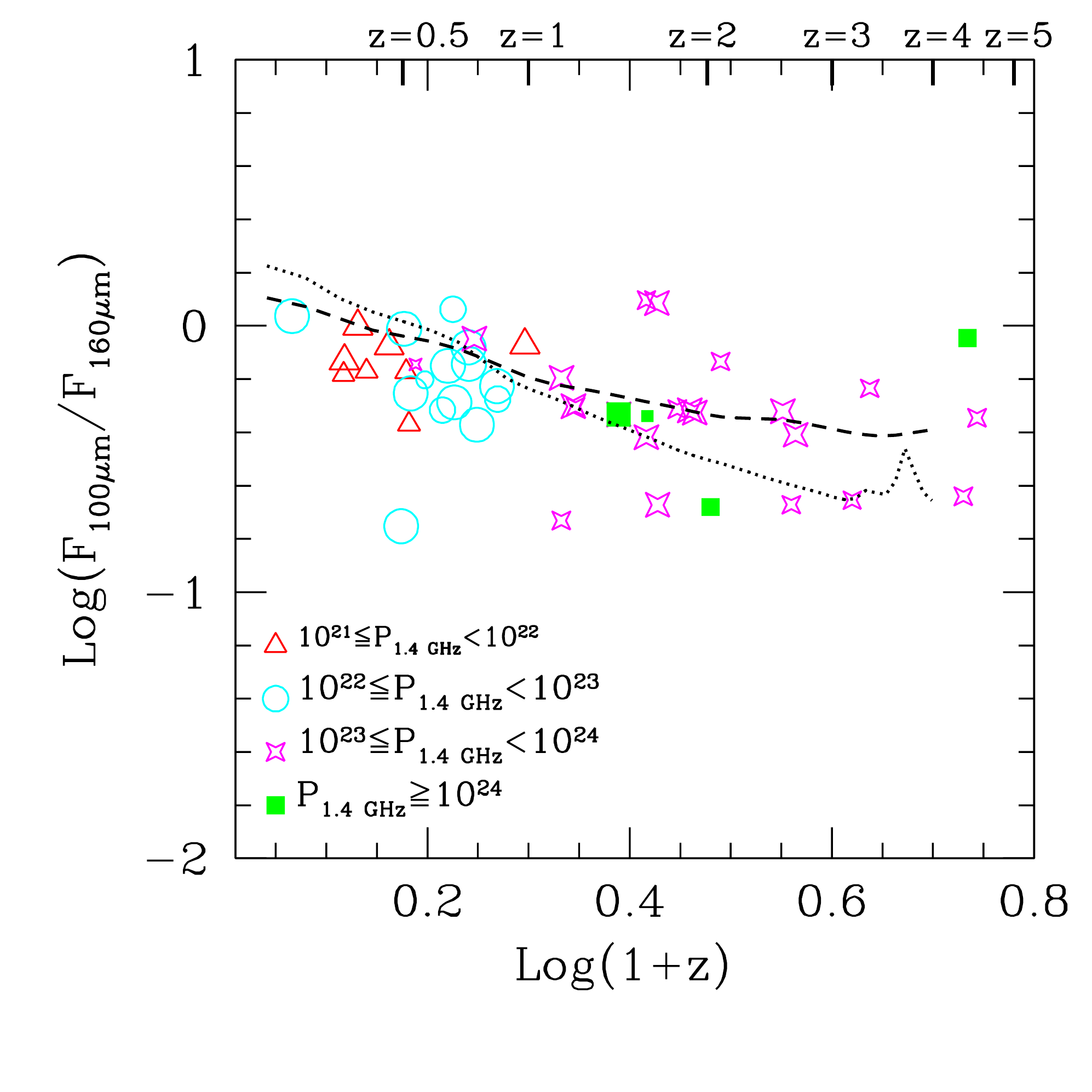}
\caption{FIR colour distribution as a function of redshift for those radio-active AGN which also emit at FIR wavelengths.  As illustrated by the plot itself, sources are colour-coded according to their radio luminosity. Small symbols are for AGN within the GOODS-S, intermediate-size ones for those within the GOODS-N and large symbols are for AGN within the Lockman Hole. The dashed line represents the SED of M82, while the dotted one the SED of Arp220. 
\label{fig:q100_160}}
\end{center}
\end{figure}

The results obtained in \S 4 have clearly shown that FIR activity is a very common process amongst radio-active AGN. In fact, up to $\simgt 70$\% of the radio AGN population is found to show up in the  {\it Herschel} maps. We have also shown that, except maybe for the most local ($z\simlt 0.5$, cfr. Figure \ref{fig:AGNvsz}) universe, the probability for a radio-active AGN to also emit at FIR wavelengths is independent of redshift. This implies that there is no preferred epoch for a  galaxy to be the site of both radio-AGN and FIR activities. Furthermore, we have found that, at high enough ($z\simgt 1$) redshifts, such a probability does not even depend on the power of radio emission from the central black hole. Radio luminosity of AGN origin only starts becoming important in the low-redshift universe, whereby the chances of finding FIR emission within the host galaxy are maximum for low-luminosity sources, while they monotonically decrease for increasing radio luminosities. Perhaps more importantly, our study has proven that there is no difference between the (radio) AGN emission associated to galaxies with ongoing FIR activity and that in those galaxies which are FIR-quiet. 

All the above conclusions strongly suggest that, at least for what concerns radio emission,  the central engine ignores the existence of FIR activity within the same host galaxy, both in terms of its cosmological evolution, of its inner properties and, except for in the very local universe, also in terms of its radiative power. In other words, FIR activity does not influence radio AGN activity and -- except for in the very local universe -- the two processes seem to proceed independently of each other. 

But what happens at the host galaxy level? Is there any difference in the properties of galaxies which only host a radio-active AGN or are the sites of both radio AGN and FIR activity?
In order to answer these questions, we have then looked into the distributions of stellar masses and ages for galaxies hosts of the AGN phenomenon only and of both AGN  and FIR activity. This has only been possible for the GOODS-N and GOODS-S fields, as the Lockman Hole does not have estimates for either masses or ages of the galaxies within the field. 

The distribution of stellar masses for galaxies in the combined GOODS-S and GOODS-N fields host of the radio-active AGN selected in our work is presented in the top-left panel of Figure \ref{fig:mass_GS}. The solid histogram refers to the whole population, while the dashed histogram indicates the sub-class of FIR emitters. The bottom-left panel shows the ratio between these two quantities. 

Two features can be clearly appreciated from these plots: the first one is that AGN hosts are very massive galaxies. Indeed, the mass distribution in both fields peaks in the range M$_* = 10^{11}-10^{11.5}$ M$_\odot$, and more than 95\% of the objects have masses larger than $10^{10.5}$ M$_\odot$. Furthermore, a non negligible fraction of these sources ($\sim 26$ \%) possesses stellar masses in the range $[10^{11.5}-10^{12}]$ M$_\odot$. \\
The second feature that can be appreciated is that there is no difference in the mass distribution of FIR-active and FIR-inactive AGN. As clearly shown in the bottom-left  panel of Figure \ref{fig:mass_GS}, the two distributions are identical. Not only that. Such a constancy is seen throughout the whole redshift range probed by our data (cfr. right-hand plot of Figure \ref{fig:mass_GS}), implying that there was never any difference between the masses of the hosts of AGN phenomena alone and the masses of the hosts of both AGN and FIR activity. This finding is somehow in contradiction with the results of Magliocchetti et al. (2014) on the COSMOS field, which indicated that FIR emitters preferred somehow slightly lower-mass galaxies (Log$_{10}$[M$_*^{\rm AGN}/$M$_\odot$]-Log$_{10}$[M$_*^{\rm AGN+FIR}/$M$_\odot]\sim$ 0.5). However, the discrepancy can be explained by recalling that COSMOS is covered by much shallower FIR observations than the two GOODS fields and therefore the information one can gather from its analysis is possibly affected by biases due to incompleteness. Indeed, only $40$\% of the radio-active AGN in COSMOS were found to have a counterpart in {\it Herschel} maps, while this percentage rises to $\simgt 70$\% in the GOODS fields.

\begin{figure*}
\begin{center}
\includegraphics[scale=0.42]{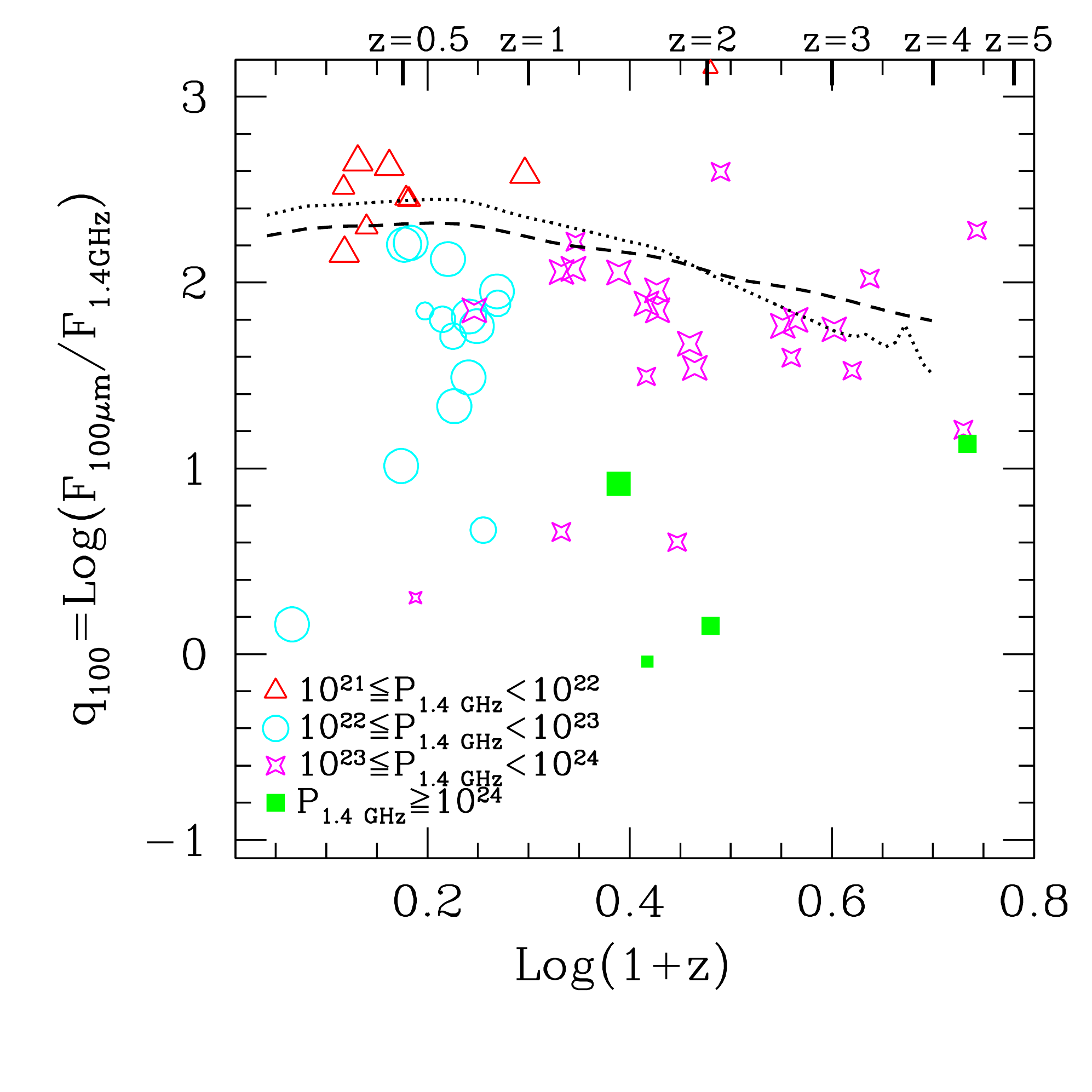}
\includegraphics[scale=0.42]{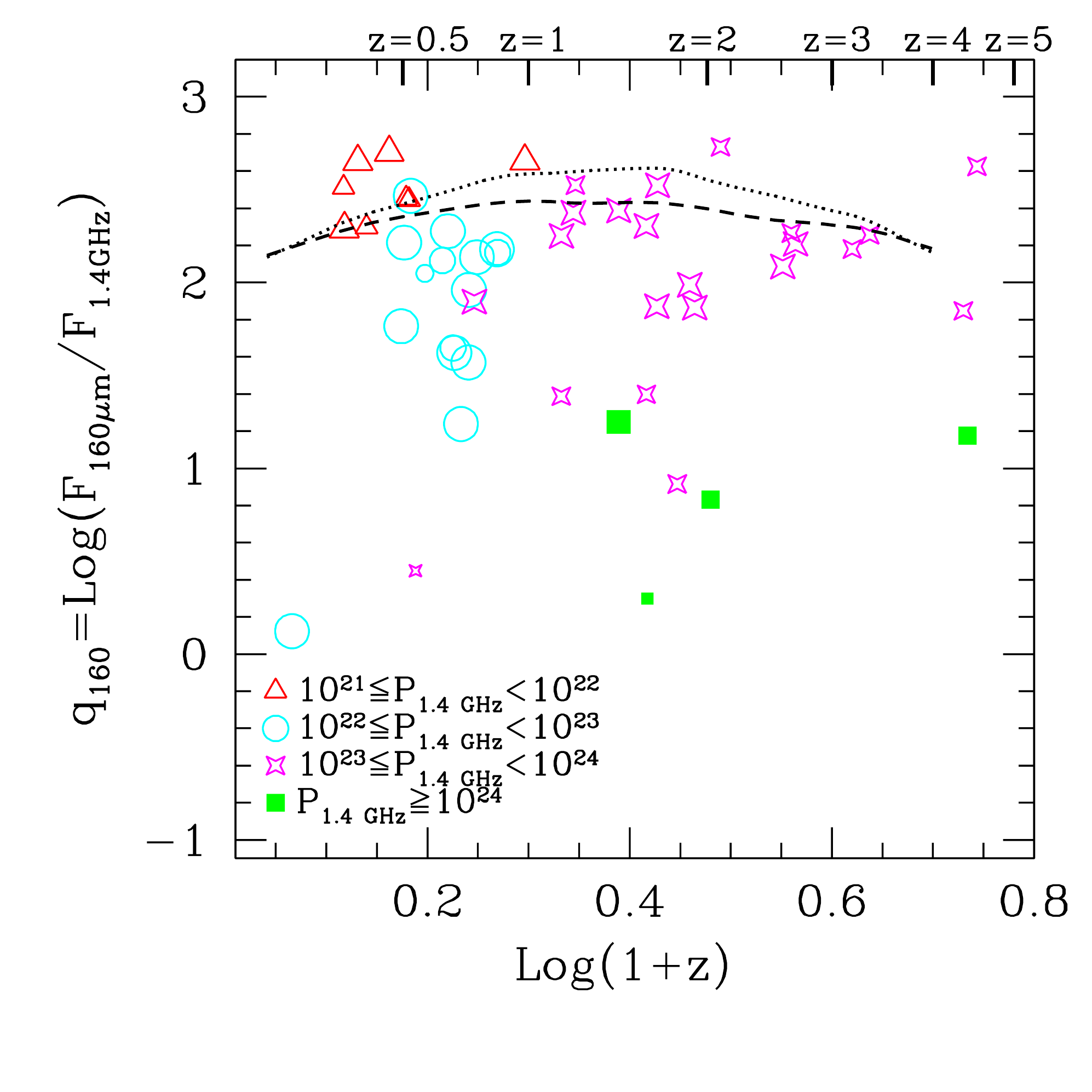}
\caption{Distribution of the quantities  q$_{100}$=Log [F$_{1.4\rm GHz}$/F$_{100\mu\rm m}$] (left-hand plot) and q$_{160}$=Log [F$_{1.4\rm GHz}$/F$_{160\mu\rm m}$] (right-hand plot) as a function of redshift for those radio-active AGN which also emit at FIR wavelengths.  Sources are colour-coded according to their radio luminosity. Small symbols are for AGN within the GOODS-S, intermediate-size ones for those within the GOODS-N and large symbols are for AGN within the Lockman Hole. The dashed lines represent the SED of M82, while the dotted ones those of Arp220. 
\label{fig:q160}}
\end{center}
\end{figure*}

Some more information on these galaxies can be obtained,  albeit only in the case of the GOODS-N, from investigations of the distribution of their ages, $\tau$. This is presented in Figure \ref{fig:age}, which shows as a solid histogram the distribution of ages for the entire AGN population, and with the dashed one that for the sub-population of FIR emitters. Although the statistical significance of the result ($\sim 1 \sigma$) is hampered by the small number of sources, Figure \ref{fig:age}  suggests that even in this case the presence of FIR emission does not alter the age properties of the galaxies under exam, at least up to about 2-3  Gyr. For galaxies older than this value, instead there seems to be a deficit of FIR emitters. 
The bulk of the distribution,  both in the presence and absence of FIR emission, is found for relatively young, 
$\tau\simlt 3$ Gyr sources. Beyond this value, the number of AGN drops dramatically, as only $\sim 15$\% of them is older than this value. 

On the basis of what found  in this section, we can then conclude that neither the age, nor the mass distribution of galaxies host of a radio AGN phenomenon vary in the presence of FIR emission. 

\section{FIR properties}

Magliocchetti et al. (2014) have proven that FIR emission from the hosts of radio AGN was entirely to be attributed to star-forming processes within the galaxy itself. 
This is also true in the present case. Figure \ref{fig:q100_160} reports the distribution of FIR, F$_{100\mu\rm m}$/F$_{160\mu \rm m}$, colours as a function of redshift and for different ranges of radio luminosity. The Figure refers to all the three considered fields, whereby small symbols are for GOODS-S, intermediate-size ones for GOODS-N and large symbols are for the Lockman Hole. The dashed and dotted lines represent the Spectral Energy Distributions (SED) of two well studied templates for star-forming galaxies such as M82 (dashed line) and Apr220 (dotted line). 

It is clear that, irrespective of cosmic age and radio luminosity, and except for very few outliers, the FIR colour distribution of radio-active AGN perfectly follows those of typical moderate-to-intense star-forming galaxies.
The radio excess  due to the presence of a radio-active AGN is only visible, and not in all cases,  in the distributions of the quantities q$_{100}$=Log [F$_{100 \mu\rm m}$/F$_{1.4 \rm GHz}$] and q$_{160}$=Log [F$_{160 \mu\rm m}$/F$_{1.4 \rm GHz}$] represented in Figure \ref{fig:q160}.

So, we have shown that FIR emission from radio-selected AGN originates from star-forming processes within the host galaxy. But how bright are these sources at FIR wavelengths? And what are their star-formation rates? 
In order to provide such answers, we have then estimated the total IR luminosities, L$_{\rm IR}$, of those radio-active AGN which showed a counterpart in {\it Herschel} maps, and subsequently calculated their star-forming rates (SFR) according to the standard relation (Kennicut 1998, which holds for a Salpeter IMF and for stellar masses in the range $\sim 0.1-100$ M$_\odot$): SFR [M$_\odot$ yr$^{-1}$]=$1.8 \cdot 10^{-10}$ L$_{\rm IR}$/L$_{\odot}$.

\begin{figure*}
\begin{center}
\includegraphics[scale=0.4]{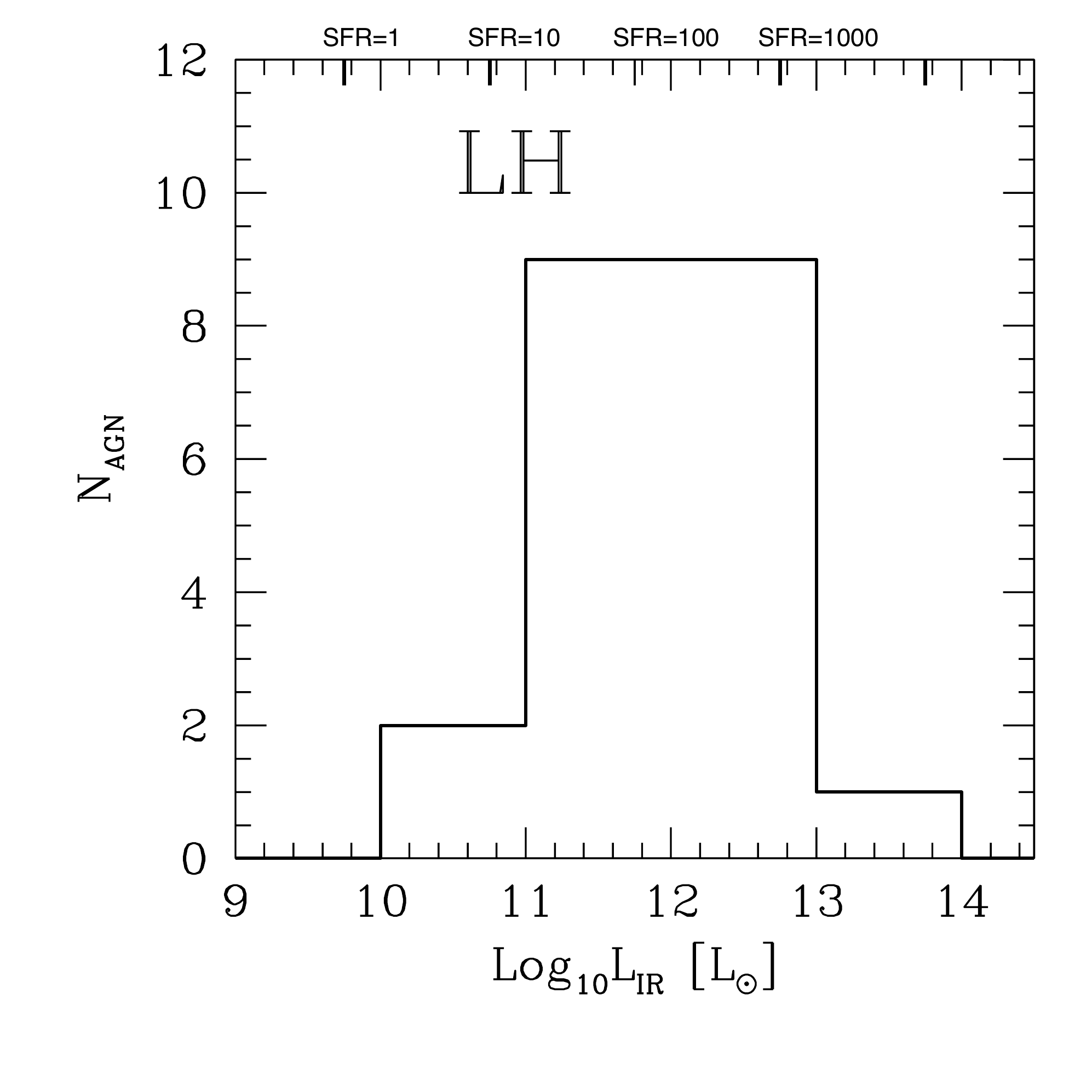}
\includegraphics[scale=0.4]{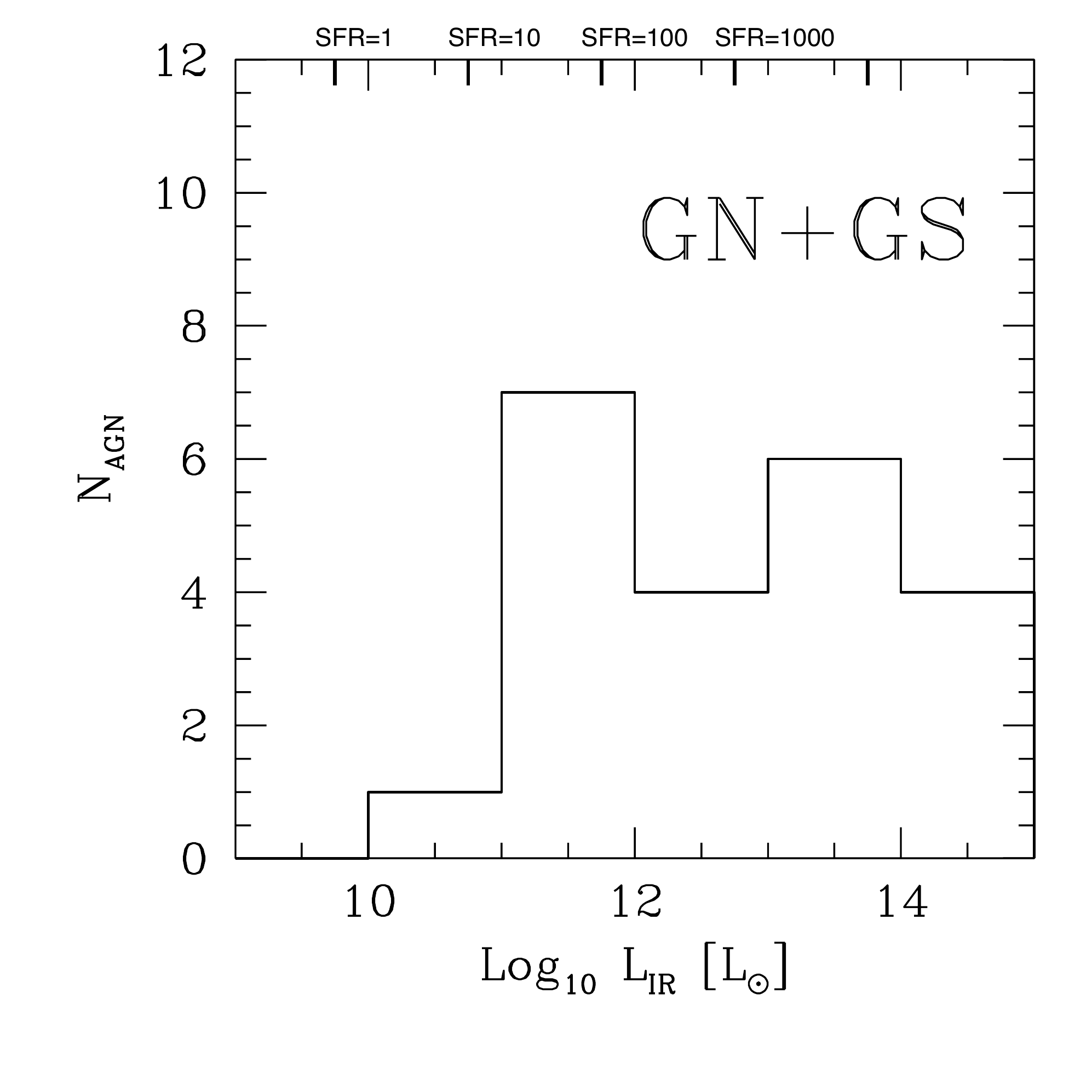}
\caption{Distribution of bolometric luminosities (L$_{\rm IR}$, bottom x axes) and star-forming rates (SFR, top x axes, expressed in M$_\odot$/yr units) for radio-active AGN selected within the Lockman Hole (left-hand panel), and the combined GOODS-S  and GOODS-N fields (right-hand panel). 
\label{fig:Lir}}
\end{center}
\end{figure*}

\begin{figure*}
\begin{center}
\includegraphics[scale=0.4]{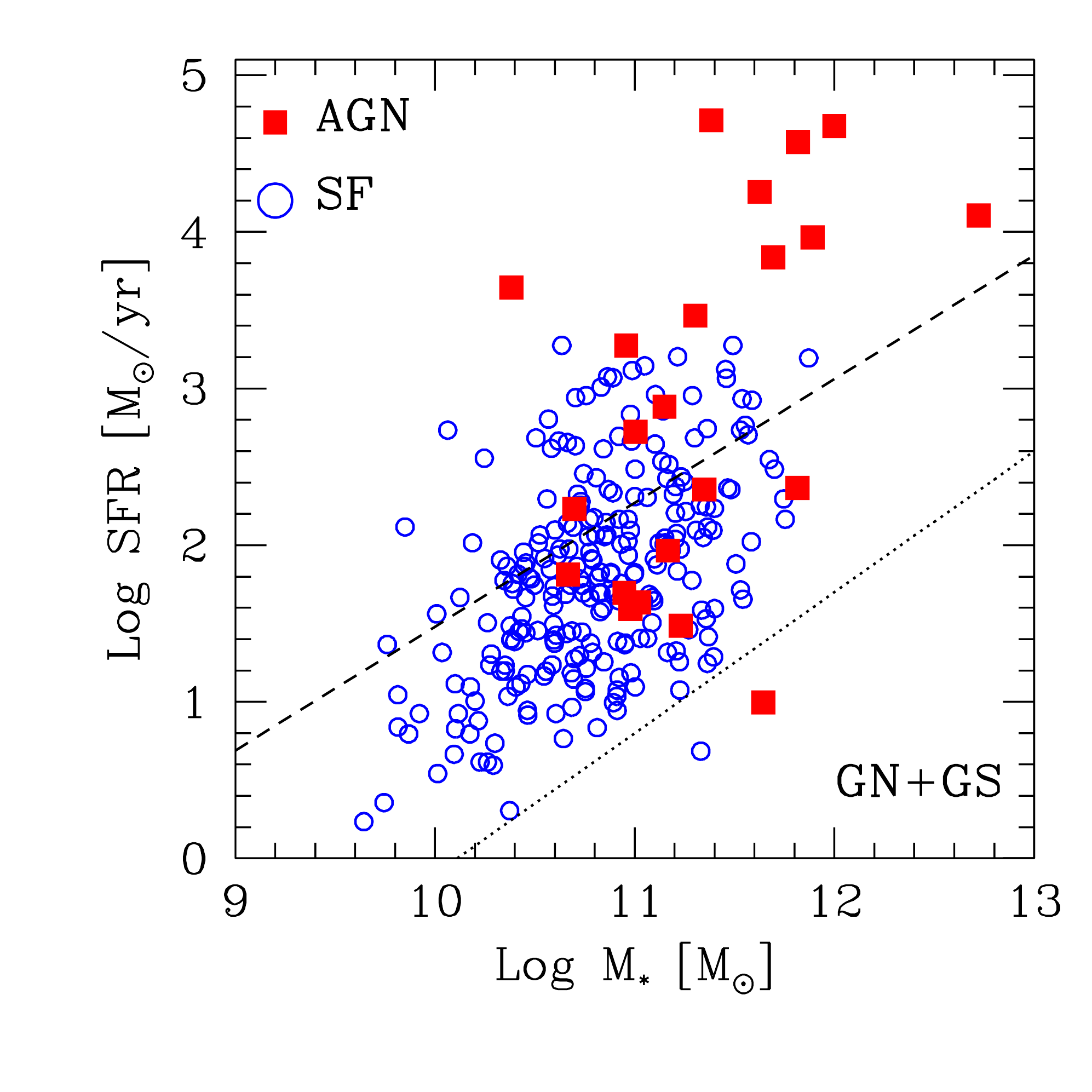}
\includegraphics[trim= 0 8 0 0, clip=true,scale=0.4]{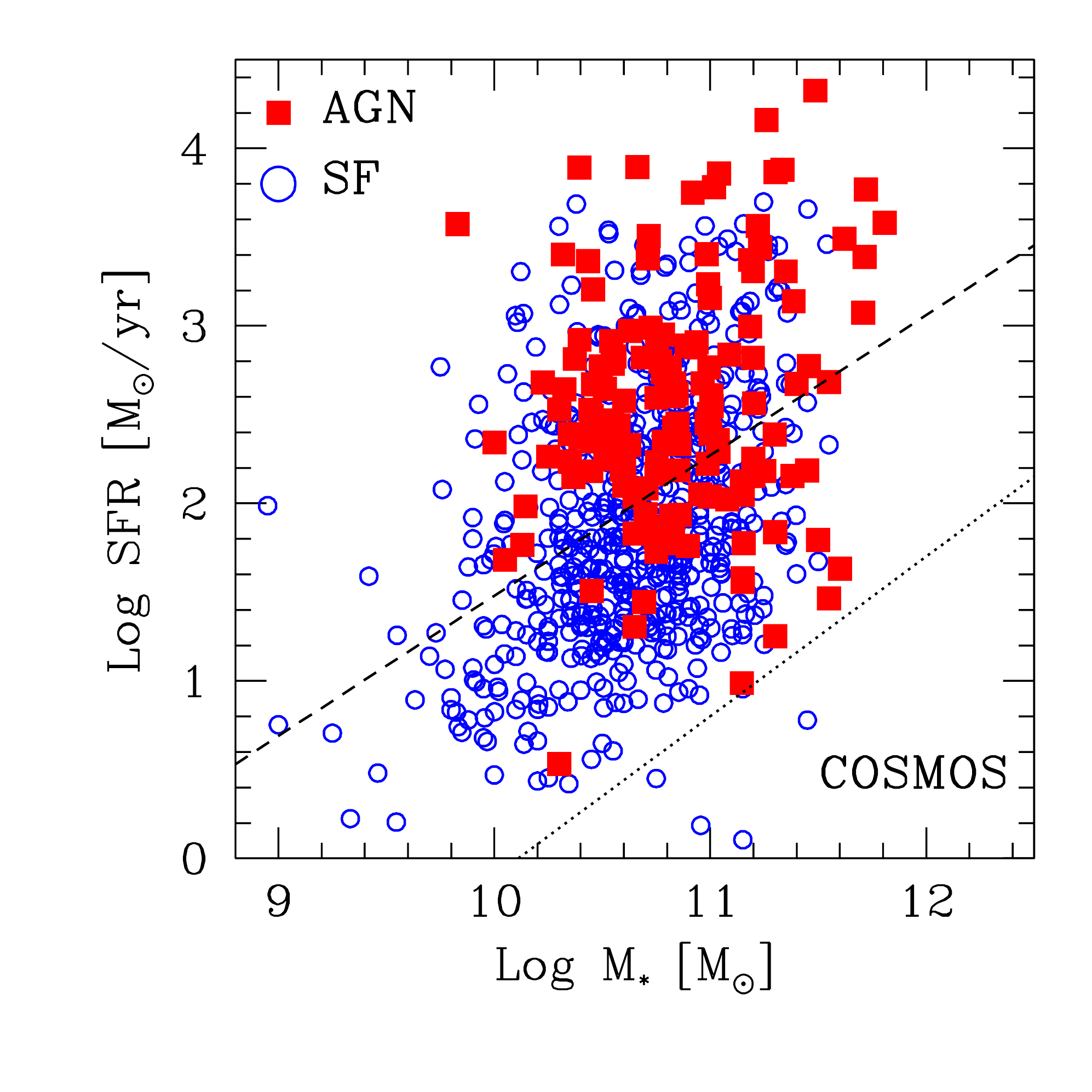}
\caption{Distributions of stellar masses vs star-formation rates for radio-selected sources at all redshifts which are also FIR-detected. Filled squares represent AGN, while open circles show star-forming galaxies. The left-hand plot refers to the combined GOODSN+GOODSS fields, while the right-hand plot illustrate the case for COSMOS. 
The dashed lines indicate the relation obtained for main sequence galaxies at z $\sim 2$ by Rodighiero et al. (2011), while the dotted lines that derived for local galaxies by Brinchmann et al. (2004) and Peng et al. (2010).
\label{fig:mass_lir}}
\end{center}
\end{figure*}

These are shown in Figure \ref{fig:Lir}, where the left-hand plot refers to the Lockman Hole,  while the right-hand one is for sources in the combined GOODS-N+GOODS-S field. The bottom x-axes report the bolometric luminosities, while SFRs are shown on the top x-axes. Note that the three objects in the combined GOODS-N+GOODS-S field endowed with bolometric luminosities above $10^{14}$ L$_\odot$ have estimated redshifts between $z\simeq 4.5$ and $z\simeq 7$. Until such redshifts are spectroscopically confirmed, caution should be taken in interpreting results connected to these galaxies.

Our data show how radio-active AGN are extremely 
luminous sources: their bolometric luminosities extend up to $\sim 10^{13}-10^{14}$ L$_\odot$, which in turn correspond to extremely high, SFR $\simgt 10^3$ M$_\odot$ yr$^{-1}$,  star forming rates. 
These results imply that galaxy hosts of radio AGN activity are indeed also the sites of extremely intense star forming emission. 
AGN activity does not seem to inhibit star formation within the host galaxy, just as FIR activity does not seem to affect radio AGN luminosity, at least in the relatively distant ($z\simgt 1$, cfr \S 4) universe. 
In other words, during most of the life-time of such sources, AGN and star formation do not seem to influence the activity of each other. Furthermore, as we learned in \S5, also the physical properties of the galaxies host of a radio-active AGN are only -- if any -- very minimally affected by the presence of a star-forming event within the same galaxy. 
Given the extremely high rate of associations which, we re-iterate, approaches $\simgt 70$\% in deep fields such as the GOODS-N, we stress that the above conclusions do not simply refer to a limited sub-sample of radio-active AGN but rather include the overwhelming majority of the considered sources: most of  radio-active AGN are associated to  intense episodes of star-formation. However, the two processes proceed independently within the same galaxy, at all redshifts except in the most local universe where, possibly intense radio activity acquires enough power to switch off the on-site star formation.

\section{The AGN contribution to galaxy evolution in radio-selected objects}

\begin{figure*}
\begin{center}
\includegraphics[scale=0.4]{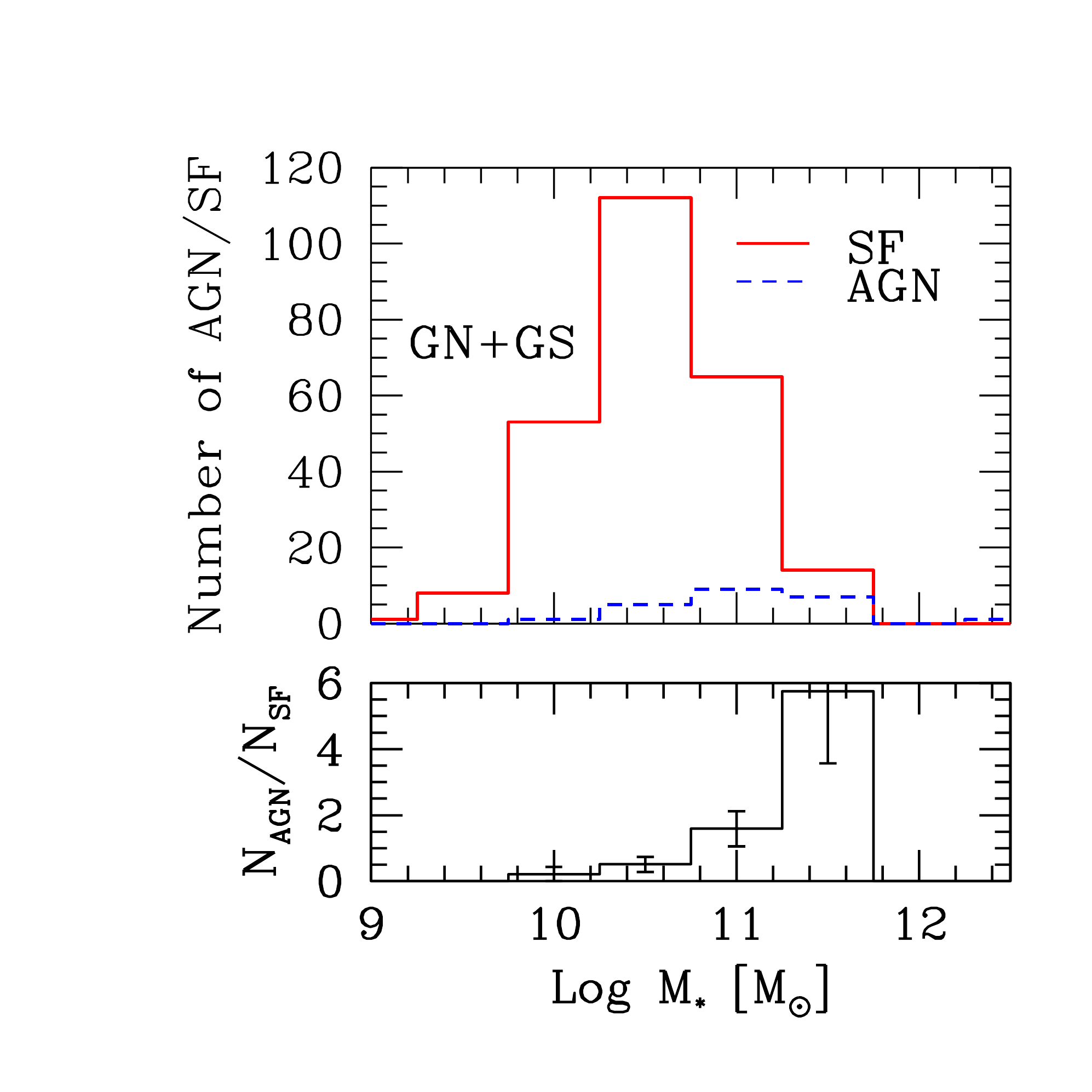}
\includegraphics[scale=0.4]{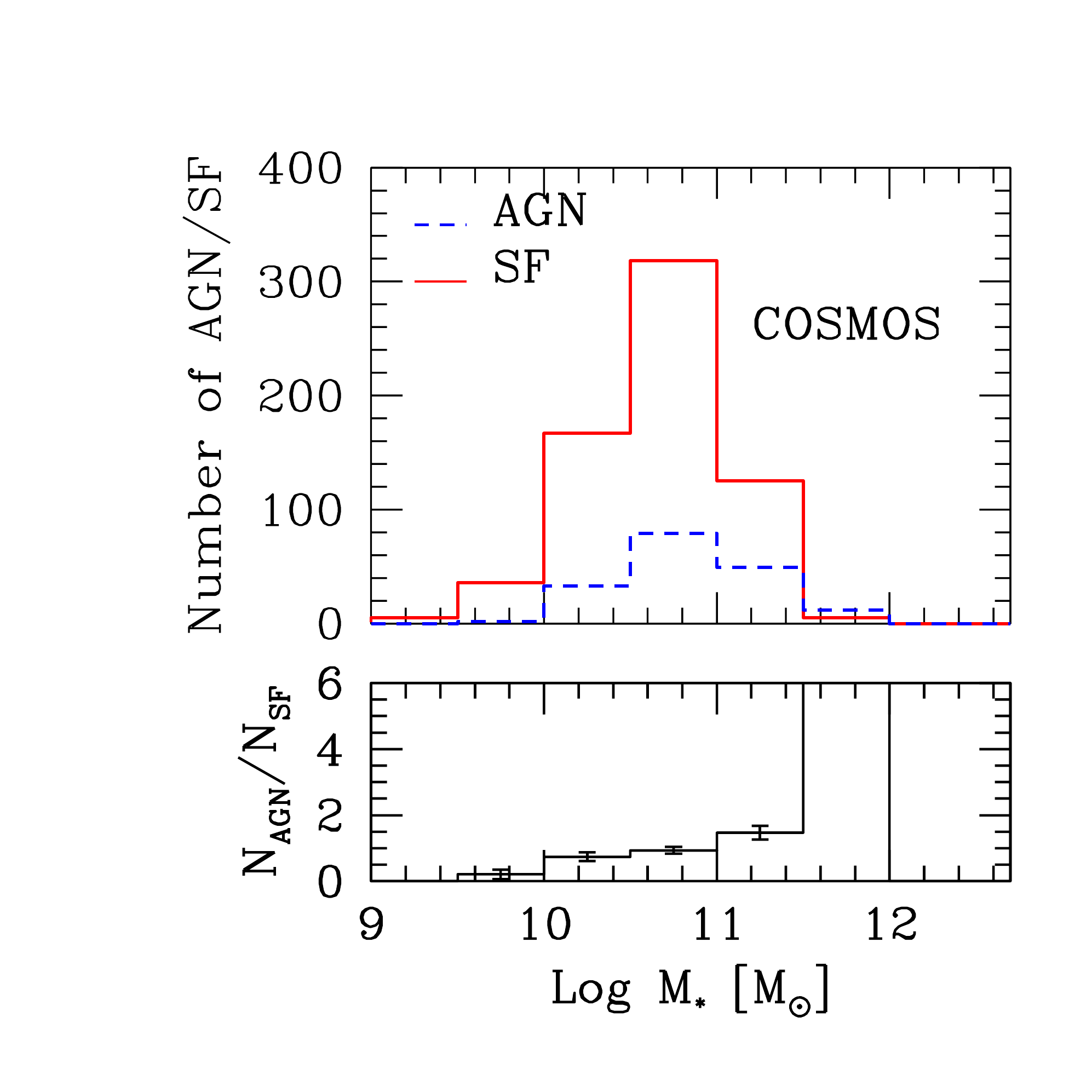}
\caption{Mass distribution for sources in the combined GOODSN and GOODSS fields (left-hand plot) and COSMOS (right-hand plot). The solid histograms represent the class of radio-selected star-forming galaxies, while the dashed ones that of radio-selected AGN. The bottom panels show the ratio between the two distributions, normalized by the total number of AGN and SF in each field. Error-bars correspond to Poissonian estimates.
\label{fig:mass_hist_SFvsAGN}}
\end{center}
\end{figure*}

In the previous sections we have shown that there is substantially no difference between the global properties of radio-selected AGN and those of the sub-population of sources which are also associated with episodes of star-formation within their hosts. These findings suggest that, at least in galaxies hosting a radio-active AGN event, star-formation is not the main trigger responsible for shaping  the formation and evolution of such sources. What we are left with is then the presence of the AGN. Is this really the main factor which influences galaxy evolution?

In order to investigate this issue in more detail, we have compared the distributions of masses and infrared luminosities/star-formation rates for the hosts of radio-selected AGN  with those of radio-selected star-forming galaxies, derived from the samples considered in the present work on the basis of their radio luminosity as explained in \S3.
This is shown in Figure \ref{fig:mass_lir}, where the filled squares refer to AGN, while the open circles to star-forming galaxies. The left-hand side plot shows the results obtained for the combined GOODSN and GOODSS fields. The right-hand side plot instead refers to COSMOS, as the Lockman Hole does not have estimates for the masses of the galaxies which reside there (cfr. \S 5). Objects in the COSMOS field belong to the sample presented by Magliocchetti et al. (2014) and have been selected in the same way as sources in the present work. In order to guide the eye, the dashed lines in both plots represent the so-called main sequence for star-forming galaxies obtained by Rodighiero et al. (2011) for z $\sim 2$ galaxies, while the dotted line shows that derived in the local universe by Brinchmann et al. (2004) and Peng et al. (2010).

Figure \ref{fig:mass_lir} clearly shows that the distribution of radio-selected sources which contain a central radio-active AGN is markedly different from that of those galaxies which do not. First of all, AGN are much more FIR-bright objects. The overwhelming majority of these sources indeed exhibits star formation rates above $\sim 100$ M$_\odot$yr$^{-1}$, while star-forming galaxies are also found down to very low IR luminosities (of the order of a few $10^{10}$ L$_\odot$). At the same time, the luminosity distribution of radio-selected AGN extends up to extremely bright luminosities, both in the case of COSMOS and in the GOODSS+GOODSS fields, while that of star-forming galaxies stops at around $10^3$ M$_\odot$yr$^{-1}$ (with slightly higher values observed in COSMOS which, we remind, includes on average brighter FIR sources due to the lower sensitivity of the {\it Herschel} observations both at 100 $\mu$m and 160 $\mu$m).

In addition, our data clearly show that star-forming galaxies are much smaller objects. Indeed, while AGN hosts have masses always  $\simgt 10^{10.5}$ M$_\odot$, star-forming galaxies can be found for stellar masses as low as $\sim 10^{9}$ M$_\odot$. On the other hand, while star-forming galaxies disappear beyond $\sim 10^{11.5}$ M$_\odot$, a non-negligible fraction of radio-selected AGN can still be found beyond that value.
This result is better illustrated in Figure \ref{fig:mass_hist_SFvsAGN} which indeed shows that, irrespective of the considered field, there is a trend for AGN to be increasingly more numerous within the population of radio-selected sources as the stellar mass of the hosts increases (cfr bottom panels of both plots).

\begin{figure*}
\begin{center}
\includegraphics[scale=0.25]{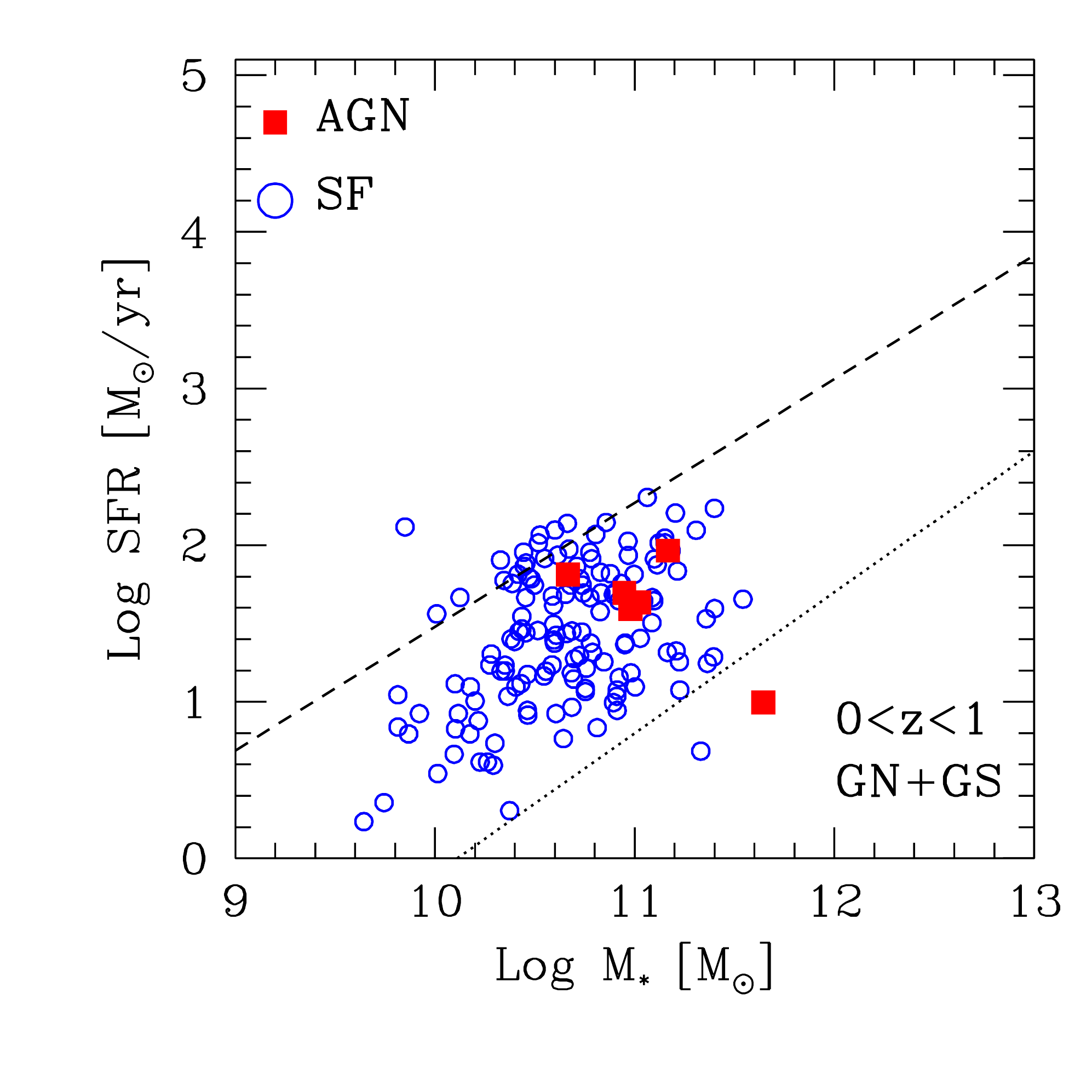}
\includegraphics[scale=0.25]{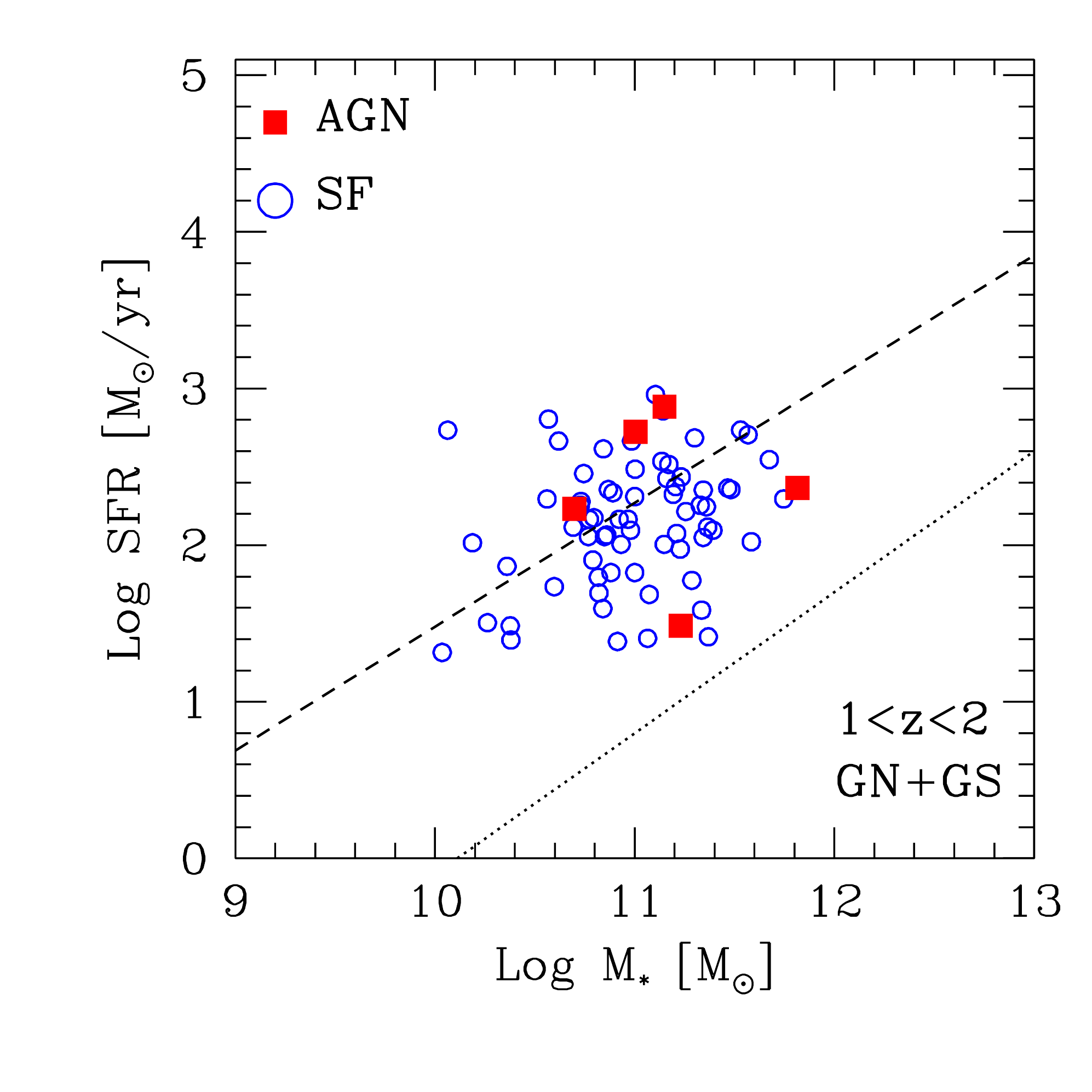}
\includegraphics[scale=0.25]{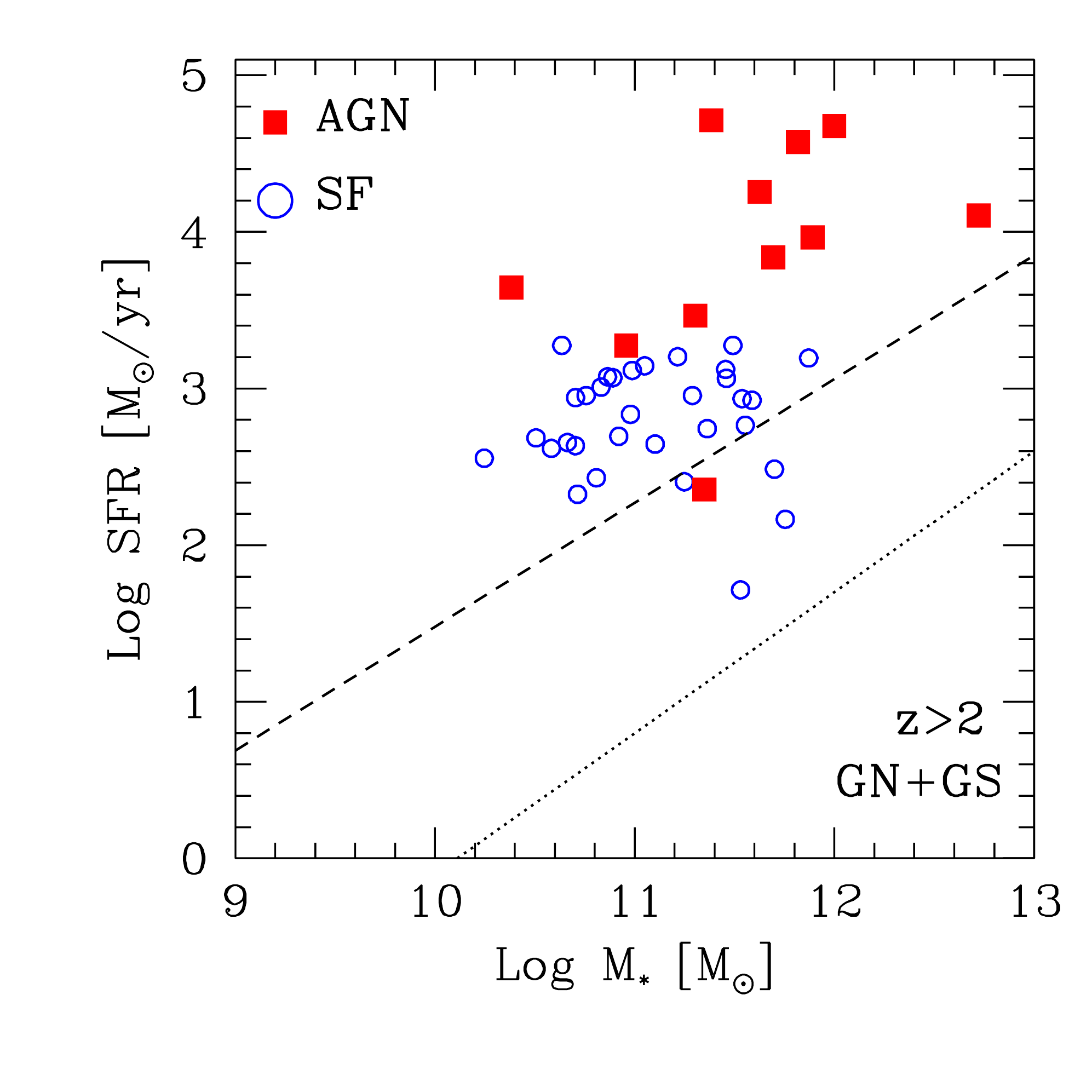}
\caption{Distributions of stellar masses vs star-formation rates for radio-selected sources which are also FIR-detected in the combined GOODDN+GOODSS fields. Filled squares represent AGN, while open circles show star-forming galaxies. Different panels refer to different redshift intervals. The dashed and dotted lines are as in Figure \ref{fig:mass_lir}.
\label{fig:mass_lir_GNandGS_z}}
\end{center}
\end{figure*}

\begin{figure*}
\begin{center}
\includegraphics[scale=0.25]{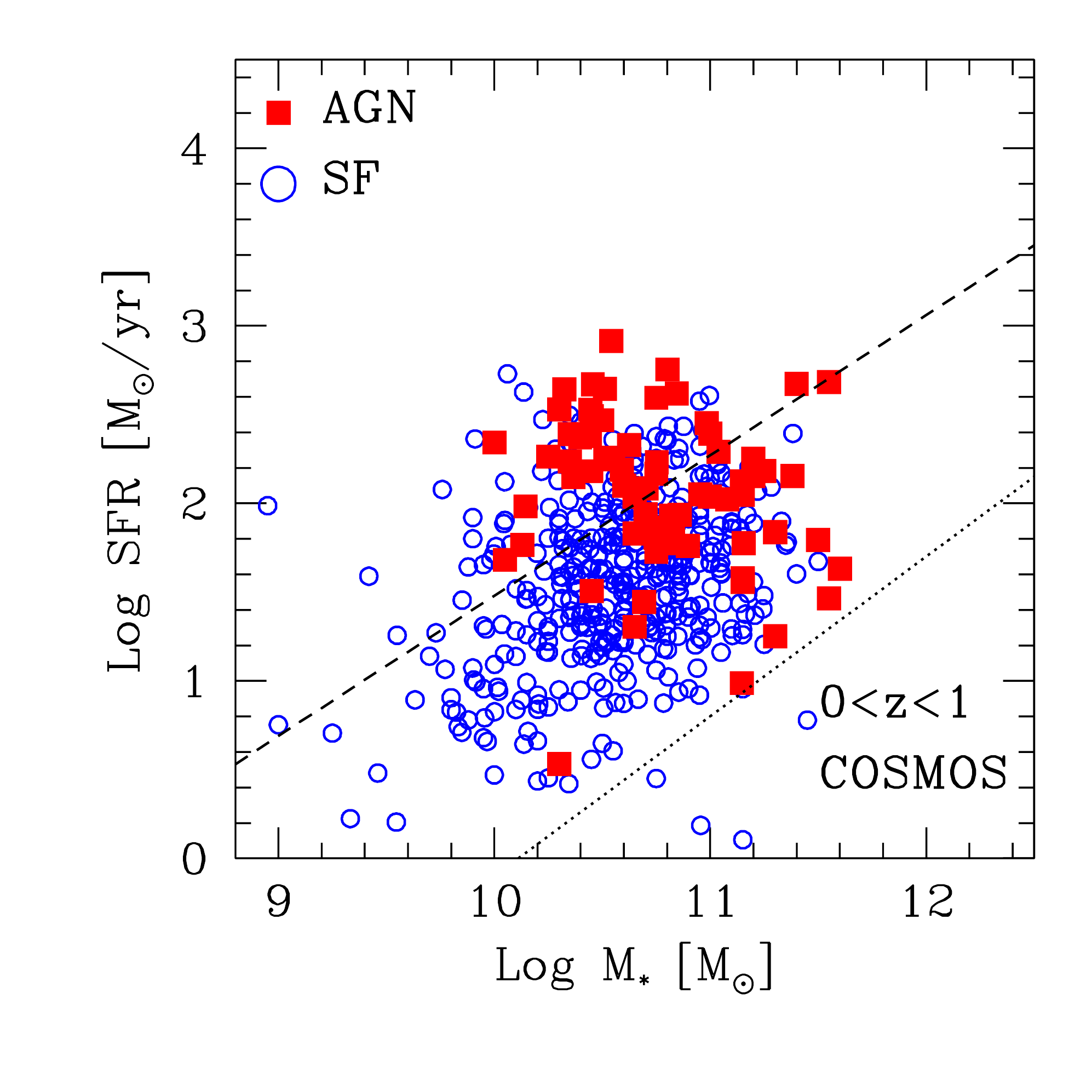}
\includegraphics[scale=0.25]{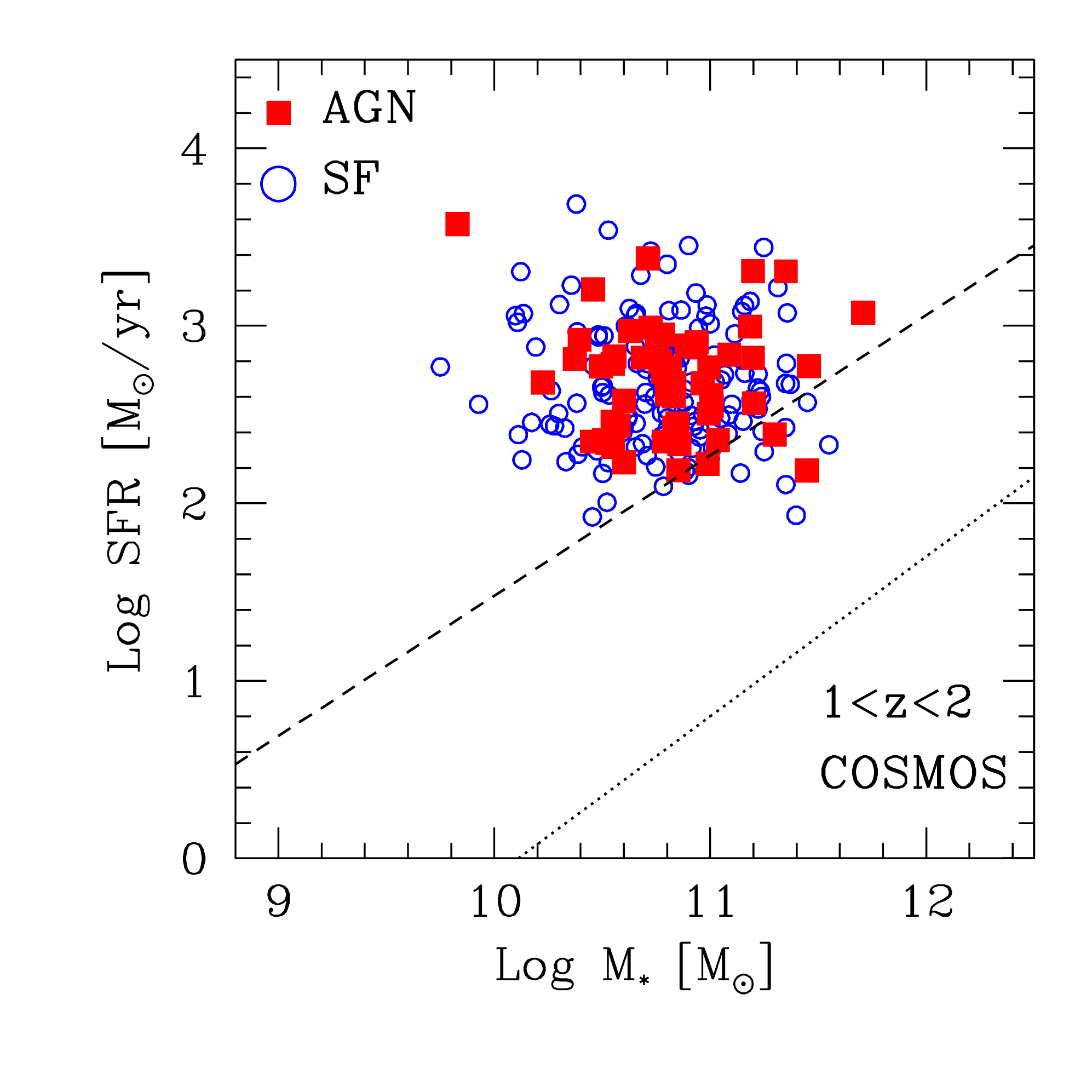}
\includegraphics[scale=0.25]{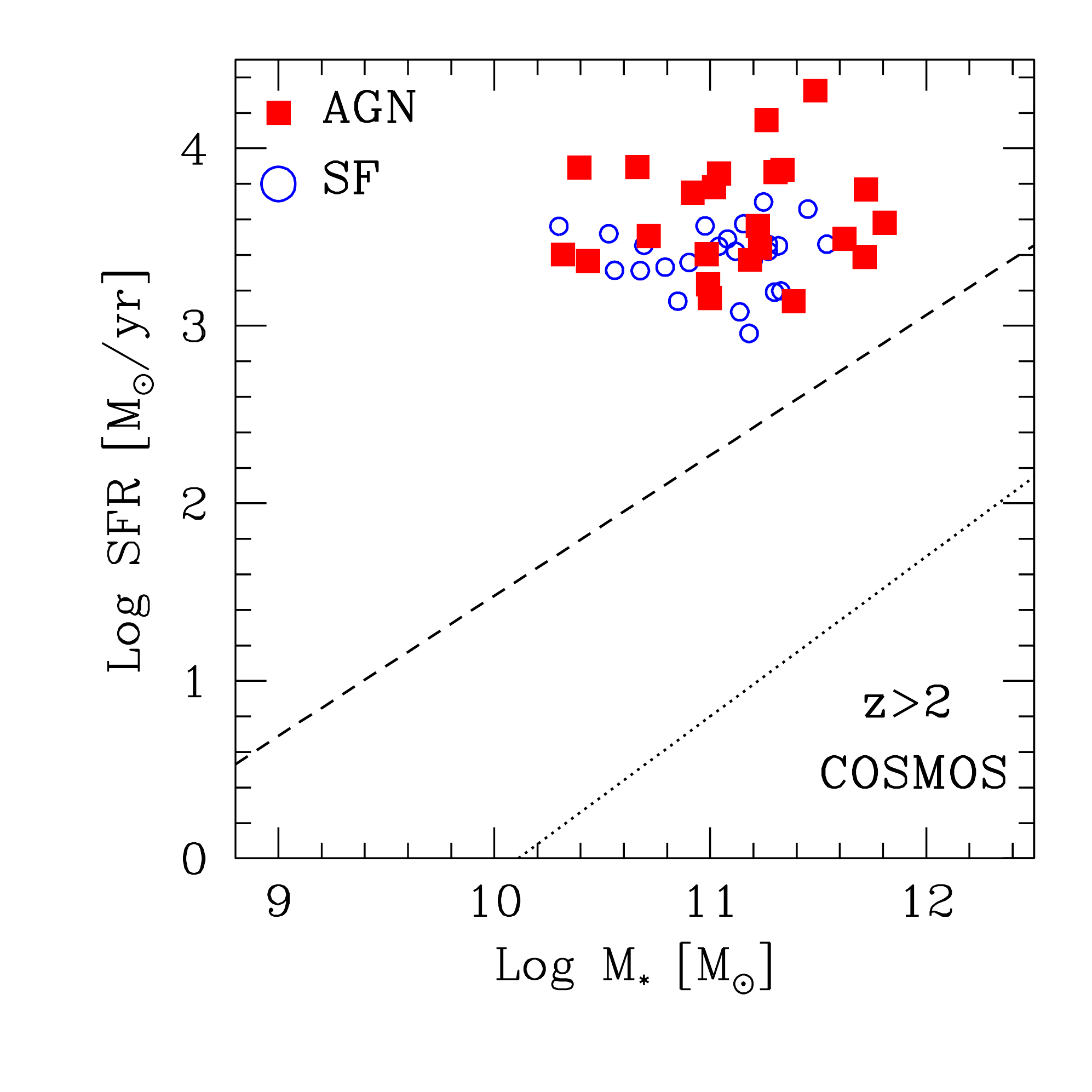}
\caption{Distributions of stellar masses vs star-formation rates for radio-selected sources which are also FIR-detected in the COSMOS field. Filled squares represent AGN, while open circles show star-forming galaxies. Different panels refer to different redshift intervals. The dashed and dotted lines are as in Figure \ref{fig:mass_lir}.
\label{fig:mass_lir_cosmos_z}}
\end{center}
\end{figure*}

So, are these differences between radio-selected AGN and star-forming galaxies inbuilt in the process of their formation or do these two populations differentiate at a certain point of their evolution? The answer can be found in the plots presented in Figures \ref{fig:mass_lir_GNandGS_z} and \ref{fig:mass_lir_cosmos_z}. These show that, a part from a number of outliers which mostly represent objects set at very high (z $\simgt 4$) redshifts and mainly lie in the combined GOODSS+GOODSN fields (see also the discussion in \S 6), the distributions of radio-selected AGN and star-forming galaxies are almost indistinguishable one from the other down to z $\sim 1$. The two populations only differentiate in the most local universe, whereby  AGN are endowed with much larger masses and luminosities than star-forming galaxies.

These results are in agreement and extend  the findings of the former sections which indicate that for redshifts above $\sim 1$ there is no difference between radio-selected AGN and the sub-population of those which are also active at FIR wavelengths. Here we have seen that, except for a few cases which mostly reside in the very early universe, at the same high redshifts there is also no difference between FIR-active, radio-selected, AGN and FIR-active, radio-selected, star-forming galaxies. 
In all cases, differences appear only in the local universe: radio-active AGN start progressively becoming FIR-quiet, the more as the higher is their radio luminosity (cfr \S 4).  
At the same time, AGN are found associated to much higher galaxy masses than star-forming galaxies.  Interestingly enough, the few AGN which still emit at FIR wavelengths in the local universe exhibit higher FIR luminosities than star-forming galaxies (cfr left-hand plots of Figures \ref{fig:mass_lir_GNandGS_z} and \ref{fig:mass_lir_cosmos_z}).



\section{Conclusions} 
With the aim of providing an analysis as exhaustive as possible of the Far Infrared Properties  (FIR) of radio-selected AGN, we have considered three very well studied fields such as the Lockman Hole, the GOODS-N and the GOODS-S. These fields are indeed provided with the deepest radio and FIR observations available up-to-date and have also been observed at a large number of other wavelengths, allowing the determination of spectroscopic and photometric redshifts for the overwhelming majority of the sources residing there.\\  
For the primary radio selection, we used data from Ibar et al. (2009), Miller et al. (2013) and Morrison et al. (2010). In all cases, these reach depths of $\sim 20 \mu$Jy at 1.4 GHz. 

In order to distinguish between radio emission due to star-formation ongoing in the host galaxy and radio emission of AGN origin, we used the method already presented in Magliocchetti et al. (2014) which is based on the radio luminosity of the sources. This marks as an AGN all those objects with 1.4 GHz luminosities brighter than a chosen threshold, which roughly corresponds to the break in the radio luminosity function of star-forming galaxies and which positively evolves with look-back time. 
The number of radio-emitting AGN found on the three considered fields with the above method is 92: 45 in the Lockman Hole, 15 in the GOODS-S and 32 in the GOODS-N.

By then relying on the very deep observations provided by the {\it Herschel} mission at 100 $\mu$m and 160 $\mu$m, we used the PEP (Lutz et al. 2011) and GOODS-{\it Herschel}  (Elbaz et al. 2011) data to look for FIR counterparts to radio-selected AGN. 
 Thanks to the extreme depth of FIR observations, we found that the large majority of radio-emitting AGN are indeed FIR emitters. The fraction of sources with ongoing FIR activity reaches values as high as $\simgt 70$\% in the case of GOODS-N and $\sim 50$\% in the Lockman Hole, where FIR observations are shallower. We found that possibly except for the very local, $z\simlt 0.5$, universe the chances for FIR emission in a radio-active AGN do not depend on the redshift of the sources, implying that FIR activity is a common event throughout the whole life-time evolution of these objects. 
 
 As shown in Magliocchetti et al. (2014), we confirm that FIR emission in radio-active AGN is entirely due to star-forming activity within the host galaxy. The bolometric luminosities of these sources are extremely high: the peak of their distribution is observed at around L$_{\rm IR}=10^{12.5}$ L$_\odot$, with AGN reaching L$_{\rm IR}$ values as high as 10$^{14}$ L$_\odot$. Converted into star formation rates (SFR), this implies sources with average SFR $\sim 10^2$ M$_\odot$yr$^{-1}$ which can even reach values up to $10^4$ M$_\odot$yr$^{-1}$. 
Recalling that the large majority of radio-active AGN are also found to be FIR emitters, we can then draw the conclusion that most of the galaxies which host a radio-active AGN are also the hosts of extremely intense star-forming activity. 
This finding confirms and extends to the radio-active case all recent results obtained on the very high star-forming activity observed in the hosts of  AGN selected in different ways and at various wavelengths (e.g. Hatziminaoglou et al. 2010; Santini et al. 2012; Rosario et al. 2013 just to mention a few).
 
Perhaps more importantly, the data show that neither the spectral slope of the radio emission, nor the radio-luminosity of the sources under examination vary in the presence of FIR emission. This is true for the spectral slope at all redshifts,
 and suggests that the central engine responsible for the AGN phenomenon is oblivious, at least at radio wavelengths, to the presence of ongoing star formation within the same galaxy. In other words, FIR activity does not seem to influence radio AGN activity.
 This is also true for radio luminosities, at least for redshifts larger than $z\sim 1$: indeed we find that, irrespective of radio luminosity, in the more distant universe virtually all radio-emitting AGN cohabit with a star-forming event. Once again, this suggests that radio luminosity, and therefore AGN activity at radio wavelengths, is not affected by the nearby star-forming processes. The situation varies in the more local universe, as there is a deficit of FIR emitters amongst bright radio AGN. This is in accordance with the general view that powerful radio-loud AGN reside in old galaxies with little or no ongoing star-forming activity. Our results then show that this lack of star-forming activity within the hosts of radio-active AGN is only true in the nearby universe, and 
possibly implies that at some late stage of their evolution, and only for weak enough FIR emission, powerful enough AGN are capable to swipe off the gas in the surrounding star-forming regions and consequently switch off the on-site star-formation.  Note that our results are in excellent agreement with those of Best et al. (2014), which also show a clear positive evolution of radio-loud AGN which are in the 'radiative-mode' (i.e. associated to cold gas and therefore to star-formation) with respect to those which are in the so-called 'jet-mode' (radiatively inefficient and generally associated to passive galaxies) between $z=0$ and $z=1$. 
Another possible interpretation of the data is provided by recent studies of the mass function of galaxies which show that there is a transition at around $z\sim 1$ between a mass function dominated by late-type galaxies to one dominated by early-type ones (e.g. Pozzetti et al. 2010; Ilbert et al. 2013).  The radio-AGN behavior might then only mirror this more general trend derived for the whole galaxy population. 
  
 Another relevant result of our analysis is that also the general properties of the galaxies host  of radio-selected AGN), such as their stellar mass or their age, do not vary in the presence of ongoing FIR activity. This means that the hosts of radio-active AGN all belong to the same population, irrespective of whether or not there is ongoing star-forming activity, i.e. that star-formation is not the dominant ingredient in shaping a galaxy which also hosts a radio-active AGN.
 
 We stress once again that the above results rely on an extremely high success rate of radio-to-FIR associations and therefore that the conclusions we can draw with the present work refer to the majority of radio-active AGN. 
For this class of sources no effect of star-forming origin is observed on AGN activity. All the considered properties both of the host galaxy and of the AGN itself remain unaltered in the presence of FIR emission, showing that star-formation does not influence (at least radio) AGN activity or shape the physical properties of the hosting galaxies. On the other hand, most of such AGN hosts (virtually the totality for redshifts $z\simgt 1$) are also found to be the sites of even extremely intense star-forming activity. This indicates  that also  star formation activity is not affected by the presence of an AGN, in all cases except for radio-luminous sources in the local, $z\simlt 1$, universe. 
We can then conclude that, except for the case of relatively local sources and  high radio luminosities, no interaction is observed between the processes of AGN emission and star-formation, as they seem to proceed independently of each other.  Note that no clear evidence for the dominance of AGN-feedback in shaping galaxy evolution has also been found by Pozzi et al. (2015), who use a backward approach capable to reproduce the galaxy K-band and the far-IR luminosity functions
  across the redshift range $0 < z < 3$.
  
  Lastly, we have shown that, except for a few cases of sources mainly residing at very high (z $\simgt 4$) redshifts,  galaxies hosting radio-selected star-forming events and those hosting radio-selected AGN are indistinguishable from each other both in terms of mass and IR luminosity (or star-forming rate)  at all redshifts larger than $\sim 1$.  Differences only start appearing in the local universe, where radio-active AGN start progressively becoming FIR-quiet, the more as the higher is their radio luminosity, and 
at the same time are found associated to much larger galaxies than those which host a star-forming event.  
 
 Why differences in the various radio-selected populations considered here only appear in the local universe is still unclear and will be investigated in a forthcoming paper.
\\
\\
\noindent
{\bf Acknowledgements}
We wish to thank the referee for his/her careful reading of the manuscript and constructive suggestions which helped shaping and improving the paper.
PACS has been developed by a consortium of institutes led by MPE
(Germany) and including UVIE
(Austria); KU Leuven, CSL, IMEC (Belgium); CEA, LAM (France); MPIA
(Germany); INAF-
IFSI/OAA/OAP/OAT, LENS, SISSA (Italy); IAC (Spain). This development
has been supported by the
funding agencies BMVIT (Austria), ESA-PRODEX (Belgium), CEA/CNES
(France), DLR (Germany),
ASI/INAF (Italy), and CICYT/MCYT (Spain).

\end{document}